\documentclass[superscriptaddress,pre,nofootinbib]{revtex4}

\usepackage{color}
\usepackage{graphicx}

\usepackage{amssymb}
\usepackage{amsmath}
\usepackage{mathrsfs}
\usepackage[hidelinks]{hyperref}
\usepackage{array}
\usepackage[T1]{fontenc}
\usepackage{epstopdf}

\graphicspath{{figs/}}
\begin{document}

\title{Rate equations, spatial moments, and concentration profiles for
mobile-immobile models with power-law and mixed waiting time distributions}

\author{Timo J. Doerries}
\affiliation{Institute of Physics \& Astronomy, University of Potsdam,
14476 Potsdam, Germany}
\author{Aleksei V. Chechkin}
\affiliation{Institute of Physics \& Astronomy, University of Potsdam,
14476 Potsdam, Germany}
\affiliation{Faculty of Pure and Applied Mathematica, Hugo Steinhaus Center,
Wroc{\l}aw University of Science and Technology, Wyspianskiego 27, 50-370
Wroc{\l}aw, Poland}
\affiliation{Akhiezer Institute for Theoretical Physics, 61108 Kharkov, Ukraine}
\author{Rina Schumer}
\affiliation{Desert Research Institute, Reno, NV 89512, USA}
\author{Ralf Metzler}
\email{rmetzler@uni-potsdam.de}
\affiliation{Institute of Physics \& Astronomy, University of Potsdam,
14476 Potsdam, Germany}
\date{\today}

\begin{abstract}
We present a framework for systems in which diffusion-advection transport of a
tracer substance in a mobile zone is interrupted by trapping in an immobile zone.
Our model unifies different model approaches based on distributed-order diffusion
equations, exciton diffusion rate models, and random walk models for multi-rate
mobile-immobile mass transport. We study various forms for the trapping time
dynamics and their effects on the tracer mass in the mobile zone. Moreover we
find the associated breakthrough curves, the tracer density at a fixed point in
space as function of time, as well as the mobile and immobile concentration
profiles and the respective moments of the transport. Specifically we derive
explicit forms for the anomalous transport dynamics and an asymptotic
power-law decay of the mobile mass for a Mittag-Leffler trapping time distribution.
In our analysis we point out that even for exponential trapping time densities
transient anomalous transport is observed.
Our results have direct applications in geophysical
contexts but also in biological, soft matter, and solid state systems.
\end{abstract}

\maketitle

\section{Introduction}

In their original formulations of Brownian motion, Einstein \cite{einstein},
Smoluchowski \cite{smoluchowski}, Sutherland \cite{sutherland} and Langevin
\cite{langevin} assumed an isotropic, homogeneous environment---and thus a
constant diffusion coefficient $D$. In the hydrodynamic limit these theories
lead to the standard diffusion equation (Fick's second law \cite{fick}) for
the probability density function (PDF) $P(\mathbf{r},t)$ to find the Brownian
particle at position $\mathbf{r}$ at time $t$ \cite{landau,vankampen}. In
more mathematical terms this means that increments of Brownian motion, on a
coarse-grained level \cite{levy}, are independent, identically distributed
random variables \cite{hughes}. Apart from the linear time dependence
$\langle\mathbf{r}^2(t)\rangle\propto Dt$ of the mean squared displacement (MSD)
the quintessential consequence of these assumptions is the Gaussian PDF of
a Brownian particle, $P(\mathbf{r},t)=(4\pi Dt)^{-d/2}\exp(-\mathbf{r}^2/
[4Dt])$ in $d$ spatial dimensions \cite{vankampen,hughes}. 
However, already in 1926 Richardson concluded from measurements of the stochastic
motion of two pilot balloons in a turbulent atmosphere that the relative
spreading, i.e., the diffusion coefficient for the relative co-ordinate
of the balloons, increases with their distance $l$, and he fitted the data
with the function $D(l)=\varepsilon l^{4/3}$ with the constant $\varepsilon
\approx0.4\mathrm{cm}^{2/3}/\mathrm{sec}$. \cite{richardson}. Batchelor,
in his work on homogeneous turbulence showed that the second moment of the
Richardson process can also be obtained by using the time-dependent scaling
$D(t)\propto t^2$ of the diffusivity instead of the Richardson-4/3-law
\cite{batchelor}. Today, anomalous diffusion with a power-law form
$\langle\mathbf{r}^2(t)\rangle\propto t^{\alpha}$ are known from a wide
range of systems. Based on the value of the anomalous diffusion exponent
one typically distinguished subdiffusion for $0<\alpha<1$ and superdiffusion
for $\alpha>1$ \cite{yossi_pt,bouchaud1,report,igor_sm,eli_pt,hoefling,diego_pt}.

\begin{figure*}
\includegraphics[width=.5\textwidth]{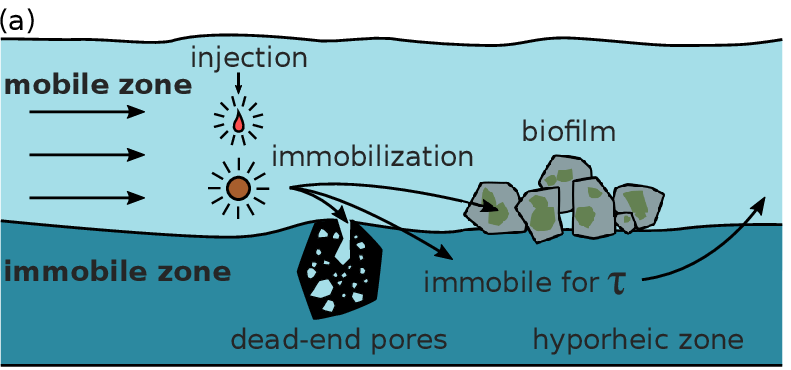}
\includegraphics[width=.375\textwidth]{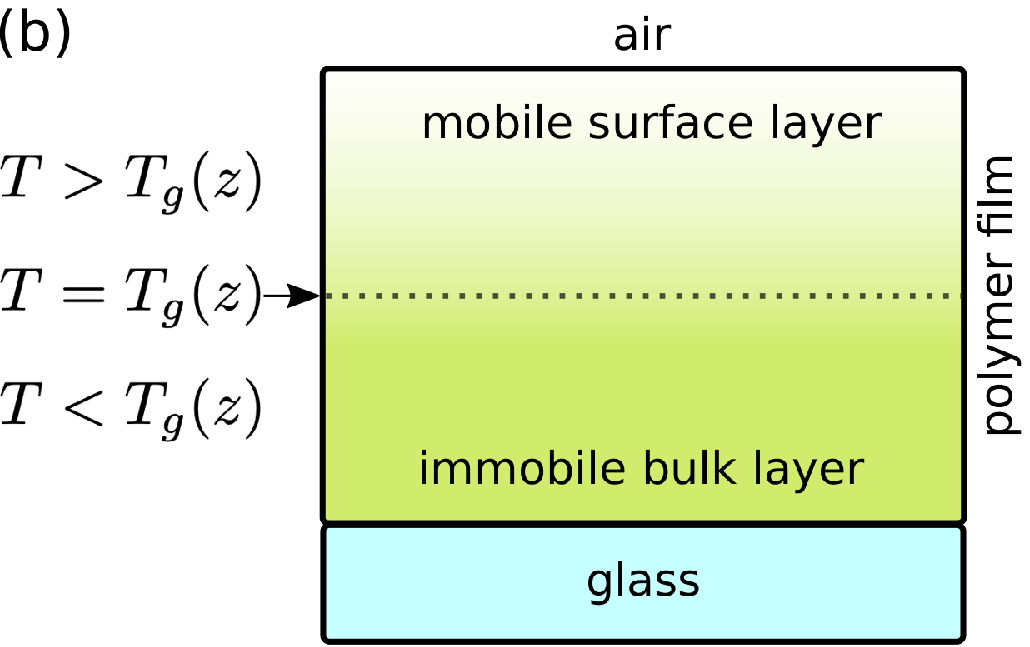}\\
\includegraphics[width=.875\textwidth]{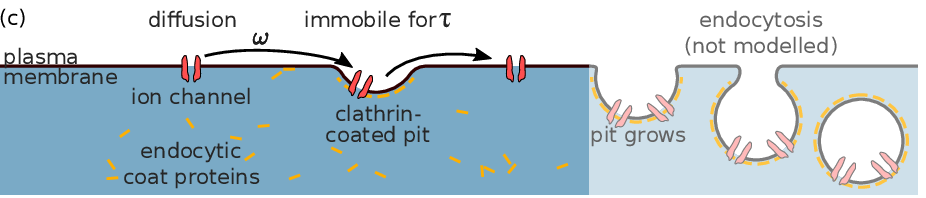}
\caption{Examples for systems with MIM zones. \textbf{(a)} Schematic of a tracer
test in a typical geophysical setting. Dye or fine particles are injected into
the mobile zone of a groundwater system or river. In our extended mobile-immobile
model (EMIM), which is based on 
\cite{kurilovich2020complex,benson2019random}, the tracers immobilize in
dead-end pores, the hyporheic zone, or biofilms for a random time $\tau$ drawn
from a probability density $\gamma(\tau)$; see Eq.~(\ref{eqCrate}) for details.
After the period $\tau$ the tracers move back into the mobile zone. 
\textbf{(b)} Schematic of a thin polymer film on a glass substrate with
depth-dependent glass transition temperature $T_g(z)$. Fluorescent dye is
immobile in the bulk layer but mobile in the surface layer. Adapted from
\cite{flier2011heterogeneous}. \textbf{(c)} Potassium channels diffuse in the
plasma membrane of human embryonic kidney cells. Endocytic coat proteins assemble
at the plasma membrane and generate clathrin-coated pits, to which the channel
binds upon encounter. The majority of channels leave the pit before the
clathrin-mediated endocytosis is completed. Adapted from
\cite{weigel2013quantifying}.}
\label{figintro}
\end{figure*}

The MSD and the particle displacement PDF are highly relevant quantities, and
they can be measured relatively straightforwardly in modern single particle
tracking experiments \cite{hoefling,lenerev}. However, they require relatively
extensive experimental setups on geological scales \cite{adams1992field}. In a
typical geophysical field experiment, as schematically depicted in
Fig.~\ref{figintro}(a), a solute or a fine particle-substance is injected into the
site and its concentration measured at selected points in space as function of time
\cite{drummond2019improving,adams1992field,goeppert2020experimental,singha2005saline,
haggerty2002powerlaw}. For Brownian tracer particles advected with a drift velocity
$\mathbf{v}$ the concentration profile has the shape $C(\mathbf{r},t)\simeq (Dt)^{
-d/2}\exp(-[\mathbf{r}-\mathbf{v}t]^2/[4Dt])$. In many geophysical experiments the
value of the PDF is measured at a given point $\mathbf{r}_0$ in space, as function
of time. This so-called breakthrough curve (BTC) at long times then shows an
asymptotic exponential decay of the tracer concentration.

In contrast to this Brownian picture, power-law tails in the time dependence of
BTCs have consistently been reported from the centimeter-scale in the laboratory
to field experiments on kilometer-scales \cite{aubeneau2014substrate,
goeppert2020experimental}. One example for such
an experiment is reported in \cite{goeppert2020experimental} based on the
injection of fluorescent dye into sinking surface water leading to a karst
aquifer under the Schwarzwasser valley, where BTCs were measured up to 7400 m
away from the injection point. In such settings, the tracer motion is
interrupted by immobilization periods, e.g. in dead end pores with negligible
flow, in which tracers are effectively trapped
\cite{coats1964deadend,haggerty2002powerlaw,gouze2008non-fickian}. 

The continuous time random walk (CTRW) is a well-established model describing
power-law tailed BTCs \cite{goeppert2020experimental,margolin2003CTRW,
dentz2003transport,brian,edery2014origins,henning}. In a CTRW a single tracer jumps
instantaneously, with variable jump lengths and waiting times drawn from respective
PDFs \cite{firststep,metzler2014anomalous}. A CTRW with scale-free, power-law
distributed waiting times and jump lengths with a finite variance was originally
introduced in the description of charge carrier motion in amorphous semiconductors
\cite{harvey} and is closely connected to the quenched energy landscape model
\cite{harvey,bouchaud,staseliprl}. When the waiting time PDF has a power-law tail of
the form $\propto t^{-1-\mu}$ with $0<\mu<1$, the associated mean waiting time
diverges and anomalous, non-Fickian diffusion arises \cite{metzler2014anomalous,
harvey,firststep}. In the quenched trap model power-law waiting time PDFs are
effected by exponential distributions of trap depths \cite{bouchaud}.
In contrast, in the model developed in \cite{mora2018brownian} a particle
undergoes Fickian diffusion which is interrupted by binding to spherical traps. All
traps have the same binding energy. Using equilibrium statistics reveals that the
densities of particles inside and outside the traps are linearly coupled with a
refilling- and an escape-rate. This yields a linear MSD with rescaled time $t\to
t/(1+\lambda)$ with a positive parameter $\lambda$ depending on the mean trapping
time and trap density. For biased transport, the case of $1<\mu<2$ with finite mean
waiting time but infinite variance still exhibits transport anomalies 
\cite{wang2020fractional}. A power-law waiting time PDF (with exponential long-time
cutoff reflecting the finiteness of the system) was indeed reconstructed from the
hydraulic conductivity field in a heterogeneous porous medium \cite{edery2014origins}.
Retention of chlorine tracer in catchments was also connected with power-law or
gamma distributed immobilization times \cite{kirchner,harveygrl}. In the CTRW
picture the PDF $P(\mathbf{r},t)$ does not distinguish between mobile and trapped
particles \cite{klablushle}.

Often, experiments in a geophysical setting yield incomplete mass recovery
\cite{adams1992field,goeppert2020experimental,singha2005saline,haggerty2002powerlaw}.
For instance, the setup of the first macrodispersion experiment (MADE-1) consisted of
an array of multilevel samplers and flow-meters to obtain the plume of bromide
injected into a heterogeneous aquifer \cite{adams1992field,boggs1992field}. The
total recovered mass monotonically decreased \cite{adams1992field}. In addition
only (or preferably) tracers that are not immobilized may be measured, because they
need to enter the detector, e.g., from a ground water spring
\cite{goeppert2020experimental,harvey2000rate,schumer2003fractal,coats1964deadend}.
In such situations it is thus desirable to have a model that separates between the
mobile and immobile particle fractions. In order to distinguish between mobile and
immobile particles within the CTRW a particle is defined to be mobile if it moves
within a preset time interval \cite{dentz2003transport}. It follows that for an
exponential waiting time all particles are mobile for preset time intervals
sufficiently longer than the characteristic waiting time \cite{dentz2003transport}. 

A modeling approach that explicitly separates the two particle fractions and
that is particularly popular in hydrology modeling, is the mobile-immobile model
(MIM) splitting the domain into mobile and immobile zones as depicted in
Fig.~\ref{figintro}(a) \cite{zhang2008moments,gao2010MIM,schumer2003fractal,
coats1964deadend,vangenuchten1976mass,haggerty1995multirate}. The description in
MIM-type models typically considers one or two spatial dimensions while
transitions between the two zones occur along an eliminated dimension. In
contrast to the CTRW, where a single concentration profile describes all
tracers, the MIM thus splits the concentration into a mobile and an immobile
concentration
\cite{dentz2003transport,haggerty1995multirate,coats1964deadend,goeppert2020experimental}.
\textcolor{black}{Including a power-law distribution of transition rates between
the zones yields power-law tailed BTCs. This model called fractal MIM
\cite{schumer2003fractal} is closely related to bi-fractional differential
equations.} Most notably in
the context of this work, the MIM has been applied successfully to geophysical
systems such as groundwater aquifers, rivers, and porous media
\cite{goeppert2020experimental,drummond2016fine,drummond2019improving,
adams1992field,haggerty2002powerlaw,harvey2000rate,coats1964deadend,
cunningham1997effects}. Understanding the motion of introduced tracers in such
systems is of high relevance to understand the dynamics of contaminants in
fresh-water sources \cite{goeppert2020experimental,drummond2019improving}.

Fig.~\ref{figintro}(a) shows a schematic of such systems. In addition to dead-end
pores, in another scenario the tracer can immobilize in streambed (benthic) biofilms
\cite{roche2016benthic,roche2019modeling,drummond2016fine,aubeneau2016biofilm}. 
Specifically, the attenuation of endocrine disruptors in a stream is attributed to
sorption and biochemical reactions in biofilms \cite{writer2012fate}. Likely,
adding wood to streams creates additional depositional areas for fine particles
and bacteria, effectively increasing the immobile capacity \cite{drummond2020fine}.
Moreover we mention the hyporheic zone, the region of near-stream aquifers
\cite{haggerty2002powerlaw} that is important for, inter alia, microplastic
retention \cite{drummond2020significance}. In addition, the hyporheic zone plays
an important role for removing organic compounds from wastewater treatment plants
that enter streams \cite{schaper2008hyporheic}. The reactivity of, e.g., metformin,
a diabetes drug, is approximately 25 times higher in the hyporheic zone as compared
to the in-stream reactivity \cite{schaper2008hyporheic}. The removal depends on
hyporheic exchange fluxes \cite{schaper2008hyporheic}. The exchange of tracers
between the mobile zone of a stream and the hyporheic zone has been studied
intensively using MIM-type approaches \cite{drummond2016fine,schumer2003fractal,
drummond2019improving}.

Importantly, applications of MIM-type models go beyond geophysical settings. We
mention that mobile and immobile zones can be found in polymer systems as shown
in Fig.~\ref{figintro}(b): A thin polymer film mounted on top of a glass support
is kept at a temperature $T$ slightly below the bulk glass transition temperature
$T_g$ \cite{flier2011heterogeneous}. Due to surface effects $T_g$ is actually a
decreasing function of the height $z$ above the glass surface. Depending on $z$
the polymer is split into an immobile bulk layer and a shallow mobile surface layer
\cite{flier2011heterogeneous,keddie1994size,yang2010glass}. Single molecule
tracking experiments of fluorescent dyes in the polymer film corroborate this
picture \cite{flier2011heterogeneous}. A second example is the transport of dye
in crystalline microporous coordination polymers, showing a pronounced splitting
into populations of fast, slow and slowest fractions \cite{liao2012heterogeneous}.

Another example with mobile and immobile zones stems from biophysics and is
shown in Fig.~\ref{figintro}(c). Here the potassium channel Kv2.1 diffuses in
the plasma membrane of a human embryonic kidney cell. Upon encountering a
clathrin-coated pit, the channels immobilize \cite{weigel2013quantifying}. A
small portion of the channels in the pit is transferred inside the cell via
clathrin-mediated endocytosis \cite{weigel2013quantifying}. The majority of
channels escape the pit and continue to diffuse. The immobilization time
statistic follows a power-law waiting time density with scaling exponent
$\mu\approx0.9$ \cite{weigel2013quantifying}.

In what follows we introduce and discuss in detail the extended mobile-immobile
model (EMIM) describing the mobile and immobile concentrations of a given tracer
substance. The dynamics is governed by a trapping time PDF of particles in the
immobile zone, \textcolor{black}{which in contrast to the MIM is not restricted
to an exponential dynamic and is well defined in in the short-time limit as
compared to the case of a power-law tailed PDF in the fractal MIM. We choose
PDFs} with and without characteristic
waiting times. Note that while the EMIM we develop here is relevant to a broad
range of systems, we will mainly use the geophysical language in what follows.
The reason is that this is the one of the most classical fields in which
MIM-type models have been applied. However, the probabilistic formulation makes
it easily accessible, and amenable for modifications, in other fields.

The paper is organized as follows. In Section \ref{secmodel} we present our EMIM
in terms of partial integrodifferential equations, we present general
expressions such as the mobile mass and transport moments, and the BTCs. We
obtain specific expressions for the observables in the EMIM and discuss possible
extensions in Section \ref{secsolutions}. In section \ref{chschumer}, we derive
from our EMIM bi-fractional models \textcolor{black}{equivalent to the fractal
MIM} and obtain exact expressions for the moments using these models. A detailed
comparison of the time evolution of the mobile mass and the BTC to experimental
observations is presented in Section \ref{chExperiments}. In Section
\ref{secconc} we draw our conclusions. In the Appendices we introduce special
functions, present details of our calculations, and show additional figures
detailing the dynamics encoded in our EMIM.

\section{The EMIM}
\label{secmodel}

We depict the motion of tracer particles in the mobile and immobile zones in a
one-dimensional two-state model, reflecting the typical situation of particles
in a riverbed (or water artery), where the coordinate $x$ measures the distance
traveled along the river. Depending on its state, a tracer either contributes
to the mobile concentration $C_m(x,t)$ or the immobile concentration $C_{im}(x,
t)$. In our model tracers are initially placed in the mobile volume 
\textcolor{black}{with mobile volume per unit length} $\theta_m$
in which their motion combines advection and diffusion, mathematically captured
by the advection-dispersion operator $L(x)=-v\partial/\partial x+D\partial^2
/\partial x^2$ with the advection velocity $v$ \cite{sokolov2006reaction}. When
entering the immobile volume $\theta_{im}$ the tracers are immobilized for a
duration $t$ drawn from the trapping time PDF $\gamma(t)$, a priori of arbitrary
shape. We name this model the extended MIM (EMIM), governed by the transport
equations
\begin{subequations}
\begin{eqnarray}
\label{eqCratea}
\frac{\partial}{\partial t}C_m(x,t)&=&-\beta\omega C_m(x,t)
+\int_0^t\gamma(t-\tau)\beta \omega C_m(x,\tau)d\tau+L(x)C_m
(x,t),\\[0.2cm]
\frac{\partial}{\partial t}C_{im}(x,t)&=&\omega C_m(x,t)-\int_0^t\gamma(t-\tau)
\omega C_m(x,\tau)d\tau.
\label{eqCrateb}
\end{eqnarray}
\label{eqCrate}
\end{subequations}
Here, $\omega$ denotes the mass transfer coefficient and $\beta=\theta_{im}/\theta_{
m}$ the capacity coefficient often used in geophysical contexts. We highlight that
this EMIM is based on the two-state, non-Markovian kinetic rate equations for
exciton trapping in semiconductors developed in \cite{kurilovich2020complex} to
which we added the advection-dispersion operator.

In this formulation $\gamma(t)$ indeed denotes the trapping time PDF. As can be seen
from relation (\ref{eqCrateb}) particles entering the immobile zone at a previous
time $t-\tau$ are released back to the mobile phase with a probability $\gamma(\tau)$.
Using the masses
\begin{equation}
M_m(t)=\theta_m\int_{-\infty}^\infty dxC_m(x,t)\mbox{ and } M_{im}(t)=\theta_{im}
\int_{-\infty}^\infty dxC_{im}(x,t),
\end{equation}
in the mobile and immobile zones, respectively, we obtain total mass conservation
\begin{equation}
\frac{d}{dt}\big[M_m(t)+M_{im}(t)\big]=0.
\label{eqMassconservation}
\end{equation}
We choose the initial condition as the sharp $\delta$-peak \textcolor{black}{
$C_m(x,0)=M_0/\theta_m\delta(x)$}
and $C_{im}(x,0)=0$, which naturally arises in typical experiments
\cite{goeppert2020experimental,drummond2016fine}. Using the Fourier-Laplace
transform we obtain from (\ref{eqCrate}) the solution 
\begin{equation}
C_m(k,s)=\frac{M_0}{s+\beta\omega(1-\gamma(s))-ikv+k^2D}
\label{eqCms}
\end{equation}
for the mobile concentration, where \textcolor{black}{
	$f(k,s)=\int_{-\infty}^\infty \int_0^\infty f(x,t)\exp(-st+ikx)dtdx$ denotes
the Fourier-Laplace transform of $f(x,t)$, which we solely mark by replacing its arguments.}
We note that this equation has been previously reported, although without the
corresponding equation in the time-domain \cite{benson2019random}. For the
immobile concentration we find
\begin{equation}
C_{im}(k,s)=\omega C_m(k,s)\frac{1-\gamma(s)}{s}.
\label{eqCimks}
\end{equation}

In this two-state approach, when modeling the exchange between the mobile and
immobile zones with single-rate first order mass transfer, exponential long-time
decay arises in the BTCs and hence it cannot describe power-law tailed BTCs
\cite{haggerty2000late,schumer2003fractal}, see also the discussion below. In
the multi-rate mass transfer model (MRMT) multiple rate coefficients are
introduced \cite{haggerty1995multirate}, in which a continuous density of rates
following a power-law distribution yields the observed BTCs with a power-law
tail \cite{schumer2003fractal,haggerty2000late,haggerty1998modeling}.  A
distribution of rates occurs in heterogeneous mixtures of layers, cylinders,
spheres, or heterogeneous porous sedimentary rock
\cite{haggerty1995multirate,berkowitz2006modelling}. Often, the cumulative
function of the trapping time PDF, $\Psi=\int_t^\infty\gamma(\tau)d
\tau=1-\int_0^t\gamma(\tau)d\tau$ \cite{klablushle}, is used for the
characterization \cite{haggerty2002powerlaw}, which can, e.g., be reconstructed
from a porous medium using X-ray microtomography \cite{gouze2008non-fickian}.
If the jump length distribution is independent from the trapping time
distribution, the total concentrations of the CTRW and MRMT approaches are
indeed equivalent \cite{dentz2003transport,schmidlin1977theory}. In
\cite{schumer2003fractal} it was shown that the choice of the power-law form
$\Psi(t)=t^{-\mu}/\Gamma(1-\mu)$ with $0<\mu<1$, yields a bi-fractional
diffusion-advection equation called fractal MIM
\cite{schumer2003fractal,sandev2015distributed}. In the long-time limit of the
fractal MIM the first and second moments of the mobile concentration scale as
$\langle x_m\rangle\simeq t^\mu$ and $\langle(x_m-\langle
x_m\rangle)^2\rangle\simeq t^{2\mu}$, respectively, and reveal superdiffusion
for $\mu>1/2$ and subdiffusion for $\mu<1/2$, while the mobile particles behave
like Brownian particles with drift in the short time limit
\cite{zhang2008moments}. The choice $\Psi(t)=t^{-\mu}/\Gamma(1 -\mu)$, however,
does not yield a finite value for $\gamma(0)$ and makes $\gamma$
non-normalizable.

To circumvent this issue, we propose the EMIM (\ref{eqCrate}) that consists of
rate equations both for the mobile and immobile concentration. The trapping time
in the immobile zone is drawn from the well-defined trapping time PDF
\textcolor{black}{ which in contrast to the MIM is not restricted to an
exponential}. Our model unifies the following approaches. First, it is an
extension of the non-Markovian rate equations used to describe excitons in
semiconductors \cite{kurilovich2020complex} to which we add an
advection-dispersion operator. A similar equation exclusively for mobile tracers
without advection was presented to describe fine particle deposition in benthic
biofilms \cite{roche2016benthic}. Second, we expand the model proposed in
\cite{maryshev2009non} in which a particle is mobile for a fixed duration and
immobile for a random time drawn from a one-sided L{\'e}vy distribution; there
effectively the total concentration is considered and no separate equations are
used for mobile and immobile particles. Third, our model corresponds to a model
used for particle tracking simulations \cite{benson2019random}. In this work it
is argued that the model incorporates waiting time PDFs, and these PDFs are
included in the Fourier-Laplace representation. Here, we derive and discuss the
corresponding rate equations as functions of time and space. Fourth, our model
contains the fractal MIM \cite{schumer2003fractal} as a special case. When
considering the total concentration, i.e., the sum of mobile and immobile
concentrations, the fractal MIM is a special case of distributed order diffusion
with a bimodal distribution of fractional orders where the first order is unity
and the second ranges between zero. Moreover, we add an advective bias term to
this formulation \cite{sandev2015distributed}.

When rewriting our rate equations in terms of the survival probability, our model
matches the MRMT model in \cite{haggerty2000late}. Another set of rate equations
involving the immobilization time as a second temporal variable can be found in
\cite{ginn2016phase,ginn2009generalization}. By choosing a Mittag-Leffler (ML)
waiting time PDF our model contains the bi-fractional solute transport models
in \cite{sandev2015distributed,schumer2003fractal} in the long-time limit,
including a power-law decay of the total mobile mass, while retaining a finite
value of the memory function in the zero-time limit, $\gamma_{\mathrm{ML}}(0)$.
From a physical perspective the accumulation of immobile particles is similar
to particles diffusing in an energy landscape scattered with energetic traps
with power-law trapping times \cite{schulz2013aging,johannes1,henning1}. We note
that while many studies focus on BTCs, some work has been reported regarding the
spatial tracer plumes \cite{adams1992field,bradley2010fractional,schumer2003fractal,
michalak2000macroscopic}.
We here address the question of where the contaminants are in space and how far they
spread on average, given a known BTC. Spatial moments of the total concentration and
their derivative, the center of mass velocity, were, inter alia, discussed in
\cite{dentz2003transport}. We here distinguish between mobile and immobile
distributions, reflecting that in some situations---including the transport
dynamics in rivers---only the mobile particles can be detected
\cite{bradley2010fractional,goeppert2020experimental}.
In \cite{zhang2008moments}, approximations for the first five moments are derived,
inter alia, for the fractal MIM including moments of the mobile plume.
Building on such concepts, from our ML waiting time PDF, we obtain explicit
expressions for the spatial moments of the mobile, immobile, and total mass.

\subsection*{General expressions}
\label{chGeneralExpressions}

We now present the central observables of our model that are calculated as
function of a general trapping time PDF.

\subsubsection{Mobile mass}

We set $k=0$ in equation (\ref{eqCms}) to arrive at the mobile mass in Laplace
space. Moreover, we set $\theta_m=1$ as a unit volume, without loss of generality.
We then obtain
\begin{equation}
M_m(s)=\frac{M_0}{s+\beta\omega[1-\gamma(s)]}.
\label{eqMs}
\end{equation}
The long-time behavior depends on the exact form of the immobilization time PDF,
in particular, on whether we have a finite or infinite mean immobilization time.
Let us first assume the general waiting time PDF $\gamma_\mathrm{f}(t)$ ("f"
denotes "finite") with a finite mean $\langle\tau\rangle$. For small Laplace
variable $s$ it can be approximated by
\textcolor{black}{$\gamma_\text{f}(s)\sim1-s\langle\tau\rangle$}, which yields
the corresponding long-time limit from (\ref{eqMs}) in terms of the constant
value
\begin{equation}
\lim_{t\to\infty}M(t)=\lim_{s\to 0}sM(s)=\frac{M_0}{1+\beta\omega\langle\tau\rangle},
\label{eqMlimfinite}
\end{equation}
which is consistent with \cite{coats1964deadend,harvey2000rate,benson2019random}
since $\beta\omega\langle\tau\rangle$ corresponds to the ratio of the time spent
in the mobile to that in the immobile zone.

For a general PDF $\gamma_\text{d}(\tau)$ ("d" denotes "divergent") with
diverging mean we consider its representation for small $s$ with $0<\mu<1$ and
$\tau_\star>0$, of the form
\textcolor{black}{$\gamma_\mathrm{d}(s)\sim1-(\tau_\star s)^{\mu}$}, where
$\tau_\star$ is a scaling factor, and plug it into the general expression
(\ref{eqMs}), where we look for the long-time limit using the Tauberian theorem
\cite{feller},
\begin{equation}
\lim_{s\to0}M(s)\sim\frac{M_0}{\beta\omega\tau_\star^\mu s^\mu},\mbox{ such that }
\lim_{t\to\infty}M(t)\sim M_0\frac{t^{\mu-1}}{\beta\omega\tau_\star^\mu\Gamma(\mu)}.
\label{eqMlimdiverge}
\end{equation}
We conclude that a waiting time PDF with diverging mean will, remarkably, yield a
long-time power-law decay of the mobile mass and thus leave no particles in the
mobile zone in the long-time limit, in contrast to a waiting time PDF with finite
mean as seen in (\ref{eqMlimfinite}).

\subsubsection{Moments}

From the PDF
\begin{equation}
\rho(k,s)=\frac{1}{M_0}(C_m(k,s)+\beta C_{im}(k,s))=\frac{1}{s}\frac{s+\beta
\omega(1-\gamma(s))}{s+\beta\omega(1-\gamma(s))-ikv+k^2D}
\label{eqrhoks}
\end{equation}
in Fourier-Laplace space we can calculate the $n$th moment in Laplace space via
\begin{equation}
\langle x^n(s)\rangle=\left.(-i)^n\frac{\partial^n}{\partial k^n}\rho(k,s)\right
|_{k=0}.
\label{eqfirstmoment}
\end{equation}
We are interested in the motion of the solute in the mobile phase, as this is
the typically accessible experimental quantity \cite{goeppert2020experimental,
adams1992field,boggs1992field,coats1964deadend,roche2016benthic}. Since
the mass in the mobile phase changes over time we consider both the unnormalized
and the normalized moments, where normalization means dividing the unnormalized moment
(denoted by "u") (\ref{eqfirstmoment}) by the mobile mass (\ref{eqMs})
\cite{chechkin2009bulk,adams1992field},
\begin{equation}
\langle x_m^n(t)\rangle=\langle x^n(t)\rangle_u M_0/M_m(t).
\end{equation}
We start with the first moment. In the unnormalized form, we have
\begin{equation}
\langle x_m(s)\rangle_u\equiv\left.-i\frac{\partial}{\partial k}\frac{C_m(k,s)}{M_0}
\right|_{k=0}=\frac{v}{(s+\beta\omega[1-\gamma(s)])^2}.
\label{eqxscaled}
\end{equation}
The short time behavior \textcolor{black}{$ \langle
x_m(s)\rangle_u=\frac{v}{(s+\beta\omega[1-\gamma(s)])^2}\sim\frac{v}{s^2}$,} of
this expression can be obtained independently of the
 trapping time PDF by using the Tauberian theorem \textcolor{black}{for $s\to
 \infty$} and $\gamma(s)\leq1$, which yields $\langle x\rangle_u(t)=vt$ for
 small $t$. This is an expected result, since essentially all mass is mobile at
 $t=0$, our initial condition. We obtain
\begin{equation}
\langle x(t)\rangle\overset{t\to0}{\sim}\langle x_m(t)\rangle_u\overset{t\to0}{
\sim}vt.
\label{eqxlimsmallt}
\end{equation}
To assess the long time behavior we need to know the specific form of the waiting
time PDF $\gamma(t)$. We will analyze the long time behavior for different cases
below.

The unnormalized second moment can be calculated analogously,
\begin{equation}
\langle x_m^2(s)\rangle_u=\left.-\frac{\partial^2}{\partial k^2}\frac{C_m(k,s)}{
M_0}\right|_{k=0}=\frac{2v^2}{(s+\beta\omega[1-\gamma(s)])^3}+\frac{2D}{(s+\beta
\omega[1-\gamma(s)])^2},
\label{eqx2scaled}
\end{equation}
the corresponding normalized form $\langle x^2(t)\rangle$ follows in the time
domain by multiplication with $M_0/M_m(t)$, equation (\ref{eqMs}). The short
time behavior
\textcolor{black}{
$\langle x^2(s)\rangle_u\sim\frac{2v^2}{s^3}+\frac{2D}{s^2},$}
of the second moment can be obtained via the Tauberian theorem
\textcolor{black}{for $s\to\infty$} and the above limit form $\gamma(s)\leq1$,
which yields $\langle x^2(t)\rangle_u\sim2Dt+v^2t^2$ at short times. Since the
mobile mass is approximately $M_0$, initially we obtain Brownian motion with
advection,
\begin{equation}
\langle x^2(t)\rangle\overset{t\to 0}{\sim}2Dt+v^2t^2,
\label{eqxxlimsmallt}
\end{equation}
a result that holds for both $\langle x^2(t)\rangle_u$ and $\langle x^2(t)\rangle$
in this $t\to0$ limit.

From the general relation (\ref{eqCimks}) between $C_m$ and $C_{im}$ and the $n$th
unnormalized moment (\ref{eqxscaled}) we obtain 
\begin{equation}
\langle x^n_{im}\rangle_u=\beta\left.(-i)^n\frac{\partial^n}{\partial k^n}\frac{
C_{im}(k,s)}{M_0}\right|_{k=0}=\beta\omega\frac{1-\gamma(s)}{s}\left.(-i)^n\frac{
\partial^n}{\partial k^n}\frac{C_{m}(k,s)}{M_0}\right|_{k=0}=\langle x^n_m\rangle
_u\beta\omega\frac{1-\gamma(s)}{s}.
\label{eqximuGeneral}
\end{equation}
This quantity describes the spreading of particles in the immobile zone, as they
progress by joining the mobile phase and getting absorbed into the immobile zone
again. In expression (\ref{eqximuGeneral}) we notice the factor $\beta$, that
appears when integrating over the immobile domain, i.e., setting $k=0$, because
the immobile domain is larger by this factor than the mobile domain. In addition
we calculate the $n$th moment of the full concentration using Eq.~(\ref{eqrhoks})
for $\rho(k,s)$,
\begin{equation}
\langle x^n\rangle=\langle x^n_m\rangle_u+\beta\langle x^n_{im}\rangle_u.
\label{eqx}
\end{equation}

\subsubsection{Breakthrough curves}
\label{chBTC}

A typical tracer experiment on the field scale records the mobile concentration
at a fixed location as function of time. The obtained statistic is called the
breakthrough curve (BTC) \cite{goeppert2020experimental,roche2019modeling,
schumer2003fractal,aubeneau2016biofilm,margolin2003CTRW,gouze2008non-fickian}. 
When comparing BTCs at different sites with different volumetric fluid discharges
$Q$, it is convenient to analyze the quantity $C\times Q/M_{\mathrm{recov}}$, with
the total recovered mass $M_{\mathrm{recov}}$ \cite{goeppert2020experimental}.
Inverse Fourier transformation of (\ref{eqCms}) yields the concentration in 
space-domain,
\begin{equation}
C_m(x,s)=\frac{\exp\left(\frac{vx}{2D}\right)}{\sqrt{v^2+4\phi(s)D}}\exp\left(
-\sqrt{v^2+4\phi(s)D}\frac{|x|}{2D}\right),
\label{eqCmxs}
\end{equation}
with $\phi(s)=s+\beta\omega(1-\gamma(s))$. Its form in time-domain requires 
an explicit input for $\phi(s)$, see below.

\section{EMIM dynamics for specific trapping PDFs}
\label{secsolutions}

We now obtain explicit forms for the characteristic observables in the EMIM dynamics
for exponential and Mittag-Leffler (ML) type trapping time density functions and
discuss possible extensions of our model.

\subsection{Exponential trapping time distribution}
\label{secClassical}

We start with the choice of an exponential distribution for the trapping time PDF,
\begin{equation}
\gamma(\tau)=\omega e^{-\omega\tau},
\label{eqgammaexp}
\end{equation}
with mean $\langle\tau\rangle= 1/\omega$. 
\textcolor{black}{The variable $\omega$ is identical to the mass transfer
coefficient from the rate equations (\ref{eqCrate}).} In the following, we
demonstrate three implications of this choice. First, when choosing an
exponential distribution in the EMIM the mobile concentration reflects one state
of a general Markovian two state model. Second, it follows immediately that
matching $\omega$ with the mass transfer coefficient from our rate equations
(\ref{eqCrate}) is not a restriction when only considering mobile tracers.
Hence, we make this choice in (\ref{eqgammaexp}). Third, we show the equivalence
of both EMIM rate equations with the choice (\ref{eqgammaexp}) in the model of
\cite{coats1964deadend}.

\begin{figure}
\includegraphics[width=\textwidth]{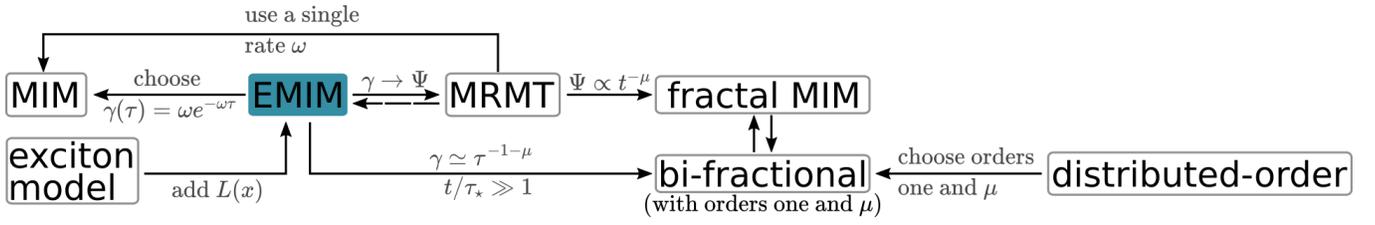}
\caption{Relation between different transport models. The EMIM follows from the exciton
model \cite{kurilovich2020complex} by adding the transport operator $L(x)$. The EMIM
contains the MIM
as a special case for $\gamma(\tau)=\omega e^{-\omega \tau}$, as shown in Section
\ref{secClassical}. Using the cumulative $\Psi$ probability function of $\gamma$,
the EMIM can be rewritten as an MRMT model, as presented in Section \ref{chschumer}.
Going the opposite direction requires $\Psi(0)=1$. If $\Psi\propto\tau^{-\mu}$, the
MRMT is equivalent to the fractal MIM \cite{schumer2003fractal}. The full
concentration follows a bi-fractional diffusion equation. The latter arises directly
from the EMIM in the long-time limit, when the waiting time PDF asymptotically
behaves like $\tau^{-1-\mu}$. The bi-fractional diffusion equation is a special
case of distributed-order diffusion where instead of a continuous distribution
the values one and $\mu$ for the diffusion exponents are chosen.}
\label{figRelationModels}
\end{figure}

To this end let us consider a general Markovian two state MIM with immobilization
rate $\omega_1$ and remobilization rate $\omega_2$ as discussed in 
\cite{kurilovich2020complex},
\begin{eqnarray}
\nonumber
\frac{\partial}{\partial t}C_m&=&-\omega_1C_m+\omega_2C_{im}+L(x)C_m,\\
\frac{\partial}{\partial t}C_{im}&=&\omega_1C_m-\omega_2C_{im},
\label{eqKurilovich}
\end{eqnarray}
where we added the advection-diffusion operator $L(x)$ to the mobile rate equation.
From (\ref{eqKurilovich}) with the initial conditions $C_{im}(x,0)=0$ and $C_m(x,0)
=M_0\delta(x)$ we obtain the formal solution \cite{kurilovich2020complex}
\begin{equation}
C_{im}(x,t)=\int_0^t\omega_1 e^{-\omega_2(t-\tau)}C_m(x,\tau)d\tau.
\label{eqCim}
\end{equation}
We insert this solution into (\ref{eqKurilovich}) to find
\begin{equation}
\frac{\partial}{\partial t}C_m(x,t)=-\omega_1C_m(x,t)+\omega_1\int_0^t\omega_2
e^{-\omega_2(t-\tau)}C_m(x,\tau)d\tau+L(x)C_m(x,t).
\label{eqKurilovich2}
\end{equation}
If we now replace $\omega_2$ with $\omega$ and $\omega_1$ with $\beta\omega$, we
recover the mobile rate equation of the EMIM (\ref{eqCratea}) with the specific
choice (\ref{eqgammaexp}). The rate equation for the immobile concentration,
\begin{equation}
\frac{\partial}{\partial t}C_{im}(x,t)=\beta\omega C_m(x,t)-\beta\int_0^t\gamma
(t-\tau)\omega C_m(x,\tau)d\tau,
\end{equation}
differs from our immobile rate equation (\ref{eqCrateb}) only by the factor $\beta$.
Note that the equivalence of the mobile concentrations suffices because $C_{im}$ is
typically not measured. We can repeat the same steps with the rate equations, that
are equivalent to the model proposed by \cite{coats1964deadend},
\begin{subequations}
\begin{eqnarray}
\frac{\partial C_m}{\partial t}+\beta\frac{\partial C_{im}}{\partial t}&=&L(x)C_m,
\label{eqCoatsa}\\
\frac{\partial C_{im}}{\partial t}&=&\omega(C_m-C_{im}),
\end{eqnarray}
\label{eqCandS}
\end{subequations}
for which we obtain 
\begin{equation}
\frac{\partial}{\partial t}C_{im}(x,t)=\omega C_m(x,t)-\int_0^t\omega e^{-\omega
(t-\tau)} C_m(x,\tau)d\tau.
\label{eqCimexp}
\end{equation}
Note specifically the equivalence with both our mobile and immobile rate equations
(\ref{eqCrate}), as can be seen from inserting (\ref{eqCimexp}) in (\ref{eqCoatsa}).
Eqs.~(\ref{eqCandS}) are first order rate equations. Therefore, we refer to the
choice $\gamma(\tau)=\omega\exp(-\omega\tau)$ as the first order model or simply
exponential model. Fig.~\ref{figRelationModels} visualizes the relation of the
EMIM to the MIM and other models.

In the long-time limit corresponding to $s\to 0$ in Laplace space, the mobile
concentration is equal to the immobile concentration, as we show in the following
calculation starting from the general relation (\ref{eqCimks}) between $C_{im}$
and $C_m$,
\begin{equation}
\lim_{s\to 0}C_{im}(k,s)=\lim_{s\to0}\omega C_m(k,s)\frac{1-\frac{\omega}{s+
\omega}}{s}=\lim_{s\to 0}\frac{\omega}{s+\omega}C_m(k,s)=\lim_{s\to 0}C_m(k,s).
\label{eqExpCmCim}
\end{equation}
Therefore, it suffices to calculate the long-time limits of the normalized moments
of the mobile concentration to obtain the long-time limits of the moments of
the immobile and full concentrations. Note that due to different mobile and immobile
volumes, the respective masses differ, which, however, does not restrict generality.

\subsubsection{Mobile mass}

Inverse Laplace transformation of the mobile mass (\ref{eqMs}) yields
\begin{equation}
	M_m(t)=\frac{M_0}{1+\beta}(1+\beta e^{-\omega(1+\beta)t}),
	\textcolor{black}{\text{~and thus }
	\lim_{t\to\infty}M_m(t)=\frac{M_0}{1+\beta}}
\label{eqClassicMass}
\end{equation}
in the long time limit, in accordance with (\ref{eqMlimfinite}) with $\langle\tau
\rangle=1/\omega$.

\subsubsection{Moments}

Using the general equation (\ref{eqxscaled}) for $\langle x_m(s)\rangle_u$ we
obtain the result for the unnormalized first moment in time domain through
inverse Laplace transformation,
\begin{eqnarray}
\langle x_m(t)\rangle_u=\frac{vt}{(1+\beta)^2}(1+\beta^2 e^{-\omega(1+\beta)t})
+v\frac{2\beta}{(1+\beta)^3\omega}(1-e^{-\omega(1+\beta)t}),
\label{eqClassicx}
\end{eqnarray}
from which we find the long-time behavior
\begin{equation}
\langle x_m(t)\rangle_u\overset{t\to \infty}{\sim}\frac{vt}{(1+\beta)^2}.
\label{eqClassicLargeTLim}
\end{equation}
We divide (\ref{eqClassicx}) by the mobile mass for normalization,
\begin{equation}
\langle x_m(t)\rangle=\frac{vt}{(1+\beta)}\frac{1+\beta^2 e^{-\omega(1+\beta)t}}{
1+\beta e^{-\omega(1+\beta)t}}+v\frac{2\beta}{(1+\beta)^2\omega}\frac{1-e^{-\omega
(1+\beta)t}}{1+\beta e^{-\omega(1+\beta)t}},
\label{eqClassicxNorm}
\end{equation}
and find the corresponding long-time behavior
\begin{equation}
\langle x_m(t)\rangle\overset{t\to \infty}{\sim}\frac{vt}{1+\beta}.
\label{eqClassicLargeTLimX}
\end{equation}
The normalization thus corresponds to rescaling time as $t\to t/(1+\beta)$. 

After Laplace inversion of (\ref{eqx2scaled}) we find the unnormalized second
moment valid at all times,
\begin{eqnarray}
\langle x_m^2(t)\rangle_u&=&\frac{2v^2}{(1+\beta)^3}\left[t^2\left(\frac{1}{2}+\frac{
\beta^3}{2}e^{-(1+\beta)\omega t}\right)+t\frac{3\beta}{(1+\beta)\omega}(1-\beta e^{
-(1+\beta)\omega t})+\frac{3(\beta^2-\beta)}{(1+\beta)^2\omega^2}\left(1-e^{-(1+\beta)
\omega t}\right)\right]\nonumber\\
&&+2D\frac{t}{(1+\beta)^2}(1+\beta^2 e^{- (1+\beta)\omega t})+2D\frac{2\beta}{(1
+\beta)^3\omega}(1-e^{-(1+\beta)\omega t}),
\label{eqxx2scaledClassic}
\end{eqnarray}
and after normalizing with $M_m(t)$, expression (\ref{eqClassicMass}), we obtain the
normalized second mobile moment
\begin{eqnarray}
\langle x_m^2(t)\rangle&=&\frac{2v^2}{(1+\beta)^2(1+\beta e^{-(1+\beta)\omega t})}
\left[t^2\left(\frac{1}{2}+\frac{\beta^3}{2}e^{-(1+\beta)\omega t}\right)+t\frac{3
\beta}{(1+\beta)\omega}(1-\beta e^{-(1+\beta)\omega t})\right.\nonumber\\
&&\left.+\frac{3(\beta^2-\beta)}{(1+\beta)^2\omega^2}(1-e^{-(1+\beta)\omega t}\right]
+2D\frac{t}{(1+\beta)}\frac{1+\beta^2 e^{-(1+\beta)\omega t}}{1+\beta e^{-(1+
\beta)\omega t}}+2D\frac{2\beta}{(1+\beta)^2\omega}\frac{1-e^{-(1+\beta)\omega t}}{
1+\beta e^{-(1+\beta)\omega t}}.
\label{eqxx2Classic}
\end{eqnarray}
In the long time limit $t\gg 1/(1+\beta)\omega$ we find from (\ref{eqxx2scaledClassic}) 
that
\begin{equation}
\langle x_m^2(t)\rangle_u\overset{t\to\infty}{\sim}\frac{v^2}{(1+\beta)^3}t^2+
2D\frac{t}{(1+\beta)^2}.
\label{eqClassicLargeT}
\end{equation}
When we account for the change of mobile mass, we obtain
\begin{equation}
\langle x_m^2(t)\rangle\overset{t\to\infty}{\sim}\frac{v^2}{(1+\beta)^2}t^2+
2D\frac{t}{1+\beta},
\label{eqClassicLargeTx}
\end{equation}
which, as for the first moment, corresponds to rescaling time $t\to t/(1+\beta)$,
see relation (\ref{eqSchumerRetardation}) in \cite{schumer2003fractal}. In fact
expression (\ref{eqClassicLargeTx}) in terms of $t/(1+\beta)$ is exactly the
expected combination of advection and diffusion of a Brownian particle in a drift
flow $v$, $v^2t^2+2Dt$.

In the long-time limit we obtain the second central moment for the classical model
in the form
\begin{equation}
\langle (x-\langle x\rangle)^2\rangle\overset{t\to\infty}{\sim}
2\frac{D}{1+\beta}t,
\label{eqxx_xLim}
\end{equation}
which grows linearly and corresponds to free Brownian motion with rescaled
time $t\to t/(1+\beta)$. These results coincide with those reported in
\cite{michalak2000macroscopic}.

\subsubsection{Breakthrough curves}

We finally calculate the long-time behavior of the mobile concentration $C_m(x,t)$, 
whose interpretation at a fixed point $x$ is that of the BTC. Starting from the
general expression (\ref{eqCmxs}) for $C_m(k,s)$ we find for small Laplace variable
$s$ that $\phi\sim s(1+\beta)$. Fourier-Laplace-inversion to the space-time-domain
yields the expected Gaussian form
\begin{equation}
	C_m(x,t)\sim \textcolor{black}{\frac{M_0}{1+\beta}\sqrt{\frac{1+\beta}{4\pi
	Dt}}}\exp\left(-\left[ x-\frac{vt}{
	1+\beta}\right]^2\textcolor{black}{\frac{1+\beta}{4D t}}\right)
\end{equation}
This result quantifies the concentration of a free Brownian particle with rescaled
time $t\to t/(1+\beta)$. Note that the immobile concentration has the same long-time
limit as shown in (\ref{eqExpCmCim}).

\subsection{Mittag-Leffler trapping time distribution}
\label{chML}

We now turn to the case when the characteristic trapping time becomes infinite and,
as explicit form, choose the generalized or two-parametric ML trapping time PDF
\cite{hilfer,sandev2018CTRW}
\begin{equation}
\gamma_\text{ML}(t)=\frac{(t/\tau_\star)^\mu}{t}E_{\mu,\mu}\left(-[t/\tau_\star]
^\mu\right),
\label{eqMLt}
\end{equation}
with $0<\mu<1$ and $\tau_\star>0$. This distribution has the power-law tail $\simeq
t^{-1-\mu}$ that indeed produces a diverging mean. We refer to the choice
(\ref{eqMLt}) as the ML model in the following. The corresponding PDF in
Laplace-domain reads \cite{gorenflo2014mittag}
\begin{equation}
\gamma_\text{ML}(s)=\frac{1}{1+(\tau_\star s)^\mu}.
\label{eqGammaML}
\end{equation}
In Section \ref{chschumer} we show that the dynamics of the total tracer
concentration in our model is a particular case of the
bi-fractional diffusion equation \cite{sandev2015distributed}, to which a transport
term is added, and the fractal model \cite{schumer2003fractal}, in the long time
limit $t\gg\tau_\star$. We note that another common choice for a PDF with a
power-law tail is the one-sided L{\'e}vy distribution
\cite{sandev2015distributed,metzler2014anomalous,kurilovich2020complex}. While
the latter is supported formally by the generalized central limit theorem
\cite{gnedenko}, its more intricate Laplace transform $\exp(-(\tau_\star s)^\mu)$
renders analytical calculations virtually impossible. As the results are expected
to be very close to those of the ML model we use the more easily tractable ML PDF
as the basis for our further study.

\subsubsection{Mobile mass decay}

For the mobile mass $M_m(s)$, see expression (\ref{eqMs}), we obtain in Appendix 
\ref{chMML} that
\begin{equation}
M_m(t)=M_0e^{-\beta \omega t}+M_0\beta\omega\tau_\star^\mu t\sum_{k=1}^\infty
(-1)^{k+1}\left(\frac{t}{\tau}\right)^{\mu k}E_{1,\mu k+2}^{k+1}(-\beta\omega
\tau_\star^\mu t),
\label{eqMm}
\end{equation}
which yields the short-time behavior
\begin{equation}
M_m(t)\sim M_0(1-\beta\omega t)+M_0\frac{\beta\omega t}{\Gamma(\mu+2)}\left(
\frac{t}{\tau_\star}\right)^\mu+O(t^{\alpha}).
\label{eqMmsmallt}
\end{equation}
Here the first term contains the initial mobile mass and immobilization with rate
$\beta\omega$. The second term contains the lowest order of the tracer
remobilization proportional to $t^{1+\mu}$. The Landau symbol $O(\cdot)$ here
represents higher order terms with $\alpha=\min(1+2\mu,2)$. Note that the series
(\ref{eqMm}) converges relatively slowly in numerical implementations.

We calculate the long-time limit of the mobile mass from its Laplace
representation.  For $t\to\infty$, corresponding to $s\to 0$, we can approximate
$\gamma_\text{ML}\sim 1-(\tau_\star s)^\mu$. We plug this form into the general
expression of $M_m(s)$, Eq.~(\ref{eqMs}), and find \textcolor{black}{
$M_m(s)\sim\frac{M_0}{ \beta\omega\tau_\star^\mu s^\mu}$ for $s\ll
1/\tau_\star$.} Via the Tauberian theorem, we obtain the result in time domain,
\begin{equation}
M_m\overset{t\to\infty}{\sim}M_0\frac{t^{\mu-1}}{\beta\omega\tau_\star^\mu\Gamma
(\mu)},
\label{eqMassLimit}
\end{equation}
in agreement with result (\ref{eqMlimdiverge}).

\subsubsection{Moments}

We calculate the long-time limits of the moments using the same approximation
$\gamma(s)_\text{ML}\sim 1-(\tau_\star s)^\mu$ for $t\to \infty$ as for the
mobile mass asymptotes. Using the general formula (\ref{eqxscaled}) for the
first unnormalized moment in Laplace space we find
\begin{equation}
\langle x_m\rangle_u=\left.i\frac{\partial}{\partial k}\frac{C_m(s)}{M_0}\right|
_{k=0}=\frac{v }{(s+\beta\omega\tau_\star^\mu s^{\mu})^2}=\frac{v }{s^2+2\beta
\omega\tau_\star^\mu s^{1+\mu}+\beta^2s^{2\mu}}\overset{s\to 0}{\sim}\frac{v}{
\beta^2\omega^2\tau_\star^{2\mu} s^{2\mu}}. 
\end{equation}
In the last step, we used that $0<\mu<1$. This corresponds to 
\begin{equation}
\langle x_m\rangle_u\overset{t\to \infty}{\sim}v\frac{t^{2\mu-1}}{\beta^2\omega
^2\tau_\star^{2\mu}\Gamma(2\mu)}.
\label{eqxUnnormalizedLimit}
\end{equation}
We now turn to the normalized first moment for large $t\to \infty$ and take the
quotient of (\ref{eqxUnnormalizedLimit}) and (\ref{eqMassLimit}), namely,
\begin{equation}
\langle x_m\rangle\overset{t\to\infty}\sim\frac{v\frac{t^{2\mu-1}}{\beta^2\omega^2
\tau_\star^{2\mu}\Gamma(2\mu)}}{\frac{t^{\mu-1}}{\beta\omega\tau_\star^\mu\Gamma(
\mu)}}=\frac{vt^\mu}{\beta\omega\tau_\star^\mu}\frac{\Gamma(\mu)}{\Gamma(2 \mu)}.
\label{eqxmlimit}
\end{equation}
The asymptote of the first moment is hence non-linear, similar to the subdiffusive
CTRW case \cite{metzler2000random}.

To obtain the asymptote of the first moment from the immobile concentration, we
start from the general relation (\ref{eqximuGeneral}) between $\langle x^n_m
\rangle_u$ and $\langle x^n_{im}\rangle_u$ for $n=1$, obtaining for $s\to0$
\begin{equation}
\langle x_{im}\rangle_u=\omega\beta v\frac{1}{(s+\beta\omega(1-\gamma(s)))^2}
\frac{1-\gamma(s)}{s}=\omega\beta v\frac{\tau_\star^\mu s^{\mu-1}}{(s+\beta
\omega\tau_\star^\mu s^\mu)^2}
\overset{s\to 0}{\sim} v\frac{ s^{-\mu-1}}{\beta\omega\tau_\star^\mu},
\label{eqximu2}
\end{equation}
which in time-domain corresponds to
\begin{equation}
\langle x_{im}\rangle\overset{t\to \infty}{\sim}\frac{vt^\mu}{\beta\omega\tau_
\star^\mu\Gamma(1+\mu)}.
\label{eqximlim2}
\end{equation}
Note that in the long-time limit all mass is immobile and we hence do not need
to normalize the moment. This result differs from $\langle x_m\rangle$ only by
the factor $\Gamma(2\mu)/\Gamma(\mu)\Gamma(1+\mu)$, which is unity for $\mu=1$
and larger than unity for $0<\mu<1$. Thus, the mobile particles travel further
on average than the immobile particles in the long-time limit, as it should be.

In what follows, we restrict ourselves to the mobile moments. Calculations of higher
immobile moments are fully analogous to the mobile moments and the first immobile
moment.

Let us turn to the second unnormalized mobile moment for $s\to 0$
\begin{equation}
\langle x^2_m\rangle_u=\left.-\frac{\partial^2}{\partial k^2}\frac{C_m(s)}{
M_0}\right|_{k=0}\overset{s\to 0}{\sim}\frac{2D}{\beta^2\omega^2\tau_\star^{
2\mu} s^{2\mu}}+\frac{2v^2}{\beta^3\omega^3\tau_\star^{3\mu}s^{3\mu}}.
\end{equation}
This corresponds in time-domain to
\begin{eqnarray}
\langle x_m^2\rangle_u\overset{t\to \infty}{\sim}2D\frac{t^{2\mu-1}}{\beta^2
\omega^2\tau_\star^{2\mu}\Gamma(2\mu)}+2v^2\frac{t^{3\mu-1}}{\beta^3\omega^3
\tau_\star^{3\mu}\Gamma(3\mu)}.
\label{eqxxUnnormalizedLimit}
\end{eqnarray}
Let us look at the normalized second moment for long $t$ and take the quotient
of (\ref{eqxxUnnormalizedLimit}) and (\ref{eqMassLimit}), namely,
\begin{equation}
\langle x^2_m\rangle\overset{t\to\infty}{\sim}\frac{2D\frac{t^{2\mu-1}}{\beta^2
\omega^2\tau_\star^{2\mu}\Gamma(2\mu)}+2v^2\frac{t^{3\mu-1}}{\beta^3\omega^3
\tau_\star^{3\mu}\Gamma(3\mu)}}{\frac{t^{\mu-1}}{\beta\omega\tau_\star^\mu
\Gamma(\mu)}}=2D\Gamma(\mu)\frac{t^\mu}{\beta\omega\tau_\star^\mu\Gamma(2\mu)}
+2v^2\Gamma(\mu)\frac{t^{2\mu}}{\beta^2\omega^2\tau_\star^{2\mu}
\Gamma(3\mu)}.
\label{eqSBMBMSDlimit2}
\end{equation}
If only mobile tracers can be observed and the waiting time PDF does not depend on
$\beta$ or $\omega$, the parameters $\beta$ and $\omega$ cannot be determined
individually, because they only appear as the product $\beta\omega$ in the
Fourier-Laplace solution (\ref{eqCms}) of $C_m(k,s)$ and all quantities derived
therefrom. Additionally in the long-time limits of the ML model the parameter
$\tau_\star$ solely appears in the product $\beta\omega\tau_\star^\mu$ and hence
cannot be determined separately. When only the long-time behavior of the mobile
tracers is known, it therefore makes sense to only consider the parameter $\beta_s'
=\beta\omega\tau_\star^\mu$. At intermediate times, the parameter $\tau_\star$
can be obtained independently from $\beta\omega$, as the mobile mass (\ref{eqMm})
shows for the ML model.

Using the asymptotes (\ref{eqxmlimit}) and (\ref{eqSBMBMSDlimit2}) of the first and 
second moment, we obtain the second central moment
\begin{equation}
\langle(x_m(t)-\langle x_m(t)\rangle)^2\rangle=2D\Gamma(\mu)\frac{t^\mu}{\beta
\omega\tau_\star^\mu\Gamma(2\mu)}+v^2\Gamma(\mu)\frac{t^{2\mu}}{(\beta\omega
\tau_\star^\mu)^2}\left(\frac{2}{\Gamma(3\mu)}-\frac{\Gamma(\mu)}{\Gamma^2(2
\mu)}\right).
\label{eqSBMBxx-xLim}
\end{equation}
The expression in the final parentheses only vanishes for $\mu=1$. In the long time
limit, the second central moment hence behaves as $t^{2\mu}$, i.e., subdiffusively
for $0<\mu<1/2$ and superdiffusively for $1/2<\mu<1$. The occurrence of a
superdiffusive behavior in a process dominated by a scale-free waiting time PDFs
is known for subdiffusive CTRW processes with drift \cite{harvey}. The phenomenon
stems from the fact that the process has a strong memory of the initial position,
its amplitude decaying only as $\simeq t^{-\mu}$. Concurrently the mobile particles
are advected, thus creating a highly asymmetric position PDF of the process. In
fact, while for a Brownian particle the ratio of standard deviation to mean
position decays as $\simeq t^{-1/2}$, for the subdiffusive particle the ratio
is asymptotically constant, reflecting the large particle spread \cite{harvey}.
This behavior is also witnessed by the slope of the concentration profiles
discussed below.

\subsubsection{Breakthrough curves}

In Appendix \ref{chConcentrationappendix} we calculate the long-time limit of the
mobile concentration using the special function of Wright type $M_\mu$ (Mainardi
function) \textcolor{black}{\cite{gorenflo2014mittag} 
\begin{equation}
M_\mu(z)=\sum_{n=0}^\infty \frac{(-z)^n}{n!\Gamma(-\mu n + (1-\mu))}.
\label{eqMwright}
\end{equation}}
\textcolor{black}{From the general equation for $C_m(x,s)$ (\ref{eqCmxs}) we find
	in the limit $s\ll [v^2/(4\beta \omega \tau^\mu_\star D)]^{1/\mu}$ for 
	$v>0$
using the Laplace inversion (\ref{eq01})}
\begin{equation}
C_m(x,t)\sim\frac{\beta\omega\tau_\star^\mu \mu}{v^2}\exp\left(\frac{v}{2D}(x-|x|)
\right)|x|t^{-1-\mu}M_\mu\left(\frac{\beta\omega\tau^\mu_\star}{v}|x|t^{-\mu}\right).
\label{eqCmlim}
\end{equation}
\textcolor{black}{For long times 
$t\gg\tau_\star(\frac{\beta\omega}{v}|x|)^{1/\mu}$ the argument of $M_\mu$ 
in (\ref{eqCmlim}) goes to zero. With the limit $M_\mu(z)\sim 1$ for $z\to 0$,
we thus have the asymptotic scaling $C_m\simeq t^{-1-\mu}$ for fixed $x$.} 
In the long-time limit, we find the immobile concentration profile 
\textcolor{black}{using the Laplace inversion (\ref{eq02})}
\begin{equation}
C_{im}(x,t)\sim\frac{\beta\omega\tau_\star^\mu}{v}\exp\left(\frac{v}{2D}(x-|x|)
\right)t^{-\mu}M_\mu\left(\frac{\beta\omega\tau^\mu_\star}{v}|x|t^{-\mu}\right).
\label{eqCimlim}
\end{equation}
Eqs.~(\ref{eqCmlim}) and (\ref{eqCimlim}) clearly show exponential cut-offs for
$x<0$, i.e., a strong suppression against the direction of the advection, as it
should be. For $x>0$ the exponential function in Eq.~(\ref{eqCimlim}) vanishes
and a cusp emerges. Conversely, at short times and fixed $x$ we find
\textcolor{black}{a Gaussian} expression of $C_m$.

\subsection{Comparison of the two EMIM cases}

When choosing an exponential trapping time distribution, our model follows the
dynamic equations (\ref{eqCandS}) corresponding to the first order mass transfer
model (\ref{eqCandS}) \cite{coats1964deadend}. In the long-time limit the mobile
and immobile concentrations are equal and the mass fraction $1/(1+\beta)$ remains
mobile. The unnormalized and normalized moments remain unchanged except for the
rescaled time $t\to t/(1+\beta)$. In the ML model, the diverging mean trapping
time leads to different mobile and immobile concentrations and a power-law decay
of the mobile mass. The first and second moments grow non-linearly and
non-quadratically in time, respectively. The second central moment shows anomalous
diffusion, i.e., subdiffusion for $0<\mu<1/2$ and superdiffusion for $1/2<\mu<1$.
All long-time limiting behaviors are summarized in Table \ref{tabkey}. In Appendix
\ref{chSim} we validate our results with particle tracking simulations.

\begin{table*}
\begin{tabular}{|l|l|l|l|l|}
\hline
& long time behavior, mobile phase & long time behavior, immobile phase \\\hline
\multicolumn{3}{|l|}{Exponential model (section \ref{secClassical})}\\\hline
\hspace{.3cm}$M_m(t)/M_0$ & $\frac{1}{1+\beta}$ (\ref{eqClassicMass})&\\\hline
\hspace{.3cm}$\langle x\rangle$ & \multicolumn{2}{c|}{$\frac{vt}{1+\beta}$,
equation (\ref{eqClassicLargeTLimX})} \\\hline
\hspace{.3cm}$\langle x^2\rangle$ & \multicolumn{2}{c|}{$2D\frac{t}{1+\beta}
+\frac{v^2 t^2}{(1+\beta)^2}$ (\ref{eqClassicLargeTx})} \\\hline
\hspace{.3cm}$\langle(x-\langle x\rangle)^2\rangle$ & \multicolumn{2}{c|}{$
2D \frac{t}{1+\beta}$ (\ref{eqxx_xLim})} \\\hline
\hspace{.3cm}$C_{m,im}(x,t)$ & \multicolumn{2}{c|}{$C_{m,im}(x,t)=\frac{M_0}{
\sqrt{\frac{2Dt}{1+\beta}}}\exp\left[-\left(x-\frac{vt}{1+\beta}\right)^2
\frac{1}{4D t/(1+\beta)}\right]$ (\ref{eqCmexplim})} \\\hline
\multicolumn{3}{|l|}{ML model (sec \ref{chML})}\\\hline
\hspace{.3cm}$M_{m,im}(t)/M_0$ & $\frac{t^{\mu-1}}{\beta\omega\tau_\star^\mu\Gamma(\mu)}$
(\ref{eqMassLimit}) & \\\hline
\hspace{.3cm}$\langle x_{m,im}\rangle$ & $v\Gamma(\mu)\frac{t^\mu}{\beta\Gamma(
\mu)}$ (\ref{eqxmlimit}) & $\frac{vt^\mu}{\beta\omega \tau_\star^\mu\Gamma(
1+\mu)}$(\ref{eqximlim2}) \\\hline
\hspace{.3cm}$\langle x_{m,im}^2\rangle$ & $ \frac{t^\mu 2D \Gamma(\mu)}{\beta
\Gamma(2\mu)}+\frac{t^{2\mu}2v^2\Gamma(\mu)}{\beta^2\omega^2\tau_\star^{2\mu}
\Gamma(3\mu)} $ (\ref{eqSBMBMSDlimit2})& \\\hline
\hspace{.3cm}$\langle(x_{m,im}-\langle x_{m,im}\rangle)^2\rangle$ & $\frac{
t^\mu2D\Gamma(\mu)}{\beta\omega\tau_\star^\mu \Gamma(2\mu)}+\frac{t^{2\mu}v^2
\Gamma(\mu)}{(\beta\omega\tau_\star^\mu)^2}\left(\frac{2}{\Gamma(3\mu)}-\frac{
\Gamma(\mu)}{\Gamma^2(2\mu)}\right)$ (\ref{eqSBMBxx-xLim}) & \\\hline
\hspace{.3cm}$C_{m,im}(x,t)$ & $\frac{\beta\omega\tau_\star^\mu \mu}{v^2}e^{
\frac{v}{2D}(x-|x|)}\frac{|x|}{t^{1+\mu}}M_\mu\left(\frac{\beta\omega\tau^\mu_
\star}{v} |x|t^{-\mu}\right)$(\ref{eqCmlim}) & $\frac{\beta\omega\tau_\star^\mu}{
vt^\mu}e^{\frac{v}{2D}(x-|x|)}M_\mu\left(\frac{\beta\omega\tau^\mu_\star}{v}|x|t^{
-\mu}\right)$ (\ref{eqCimlim})\\\hline
\end{tabular}
\caption{Main long-time behavior of the mobile immobile moments with Section and
Equation numbers. At short times, regardless of the trapping time PDF, we have
$\langle x_m\rangle\approx vt$ and $\langle x_m^2\rangle\approx v^2t^2+2Dt$, due
to $C_m(x,0)=M_0\delta(x)$ and $C_{im}(x,0)=0$.}
\label{tabkey}
\end{table*}

\subsection{Tempered power-law and composite models}
\label{chextensions}

The ML model features a diverging characteristic trapping time. While in many cases
such models reveal adequate descriptions (e.g., in \cite{kirchner,harveygrl} in
which fits with a gamma function reveal a cutoff at the very end of the experimental
window) in other cases experiments explore time ranges in which the finiteness of
the system becomes significant. A finite system size implies a finite number of
locations, e.g., pores, where the tracers can immobilize. This implies that a
finite waiting time exists, which has been measured, e.g., for dye dispersion in a
saturated sand pack \cite{berkowitz2006modelling}. A typical approach is to introduce
an exponential cut-off in the power-law waiting time PDF of the form
\cite{berkowitz2006modelling,goeppert2020experimental,edery2014origins,
dentz2003transport,roche2016benthic}
\begin{equation}
\gamma_t(t)=\frac{\exp(-t/\tau_t)}{\gamma(s=1/\tau_t)}\gamma(t),
\label{eqgammat}
\end{equation}
with the characteristic crossover time $\tau_t>0$. An interesting case is reported
in \cite{aubeneau2016biofilm} for which $\tau_t$ increases with biofilm growth. In
Laplace space we find 
\begin{equation}
\gamma_t(s)=\frac{\gamma(s+1/\tau_t)}{\gamma(s=1/\tau_t)}.
\label{eqgts}
\end{equation}
If we choose the ML model as a special example, the associated tempered PDF has
the characteristic waiting time
\begin{equation}
\langle t\rangle=\frac{\tau_\star^\mu\mu}{\tau_t^{\mu-1}(1+(\tau_\star/\tau_t)^{\mu})}.
\label{eqmwttruncated}
\end{equation}
Together with the general limit (\ref{eqMlimfinite}) of $M_m(t)$ we find
\begin{equation}
\lim_{t\to\infty}M_m(t)=\frac{M_0}{1+\beta\omega\frac{\displaystyle\tau_\star^\mu\mu}{
\displaystyle\tau_t^{\mu-1}(1+\tau_\star^\mu\tau_t^{-\mu})}}.
\label{eqMmtruncated}
\end{equation}
The assumption $\tau_t\gg\tau_\star$ appears reasonable, therefore the short time
expansion of the mobile mass coincides with the untempered ML model (\ref{eqMmsmallt}).

We now calculate an estimation of $M_m(t)$ for $\tau_\star\ll t\ll\tau_t$ using
$\gamma_t$ (\ref{eqgts}) and the general formula for $M_m$ (\ref{eqMs}) in Laplace
space,
\begin{equation}
M_m(s')=\frac{M_0}{\displaystyle s'-\frac{1}{\tau_t}+\beta\omega\left(1-\frac{1}{1+(\tau_\star s')^\mu}		
-\frac{\tau_\star^\mu\tau_t^{-\mu}}{1+(\tau_\star s')^\mu}\right)},
\label{eqmmsp}
\end{equation}
where we define $s'=s+1/\tau_t$. This definition allows us to analyze (\ref{eqmmsp})
for small $s'$
\begin{eqnarray}
M_m(s')&\overset{s'\to 0}\sim&
\frac{M_0}{\displaystyle \beta\omega\left[(\tau_\star s')^\mu		
(1+\tau_\star^\mu\tau_t^{-\mu})-\tau_\star^\mu\tau_t^{-\mu}\right]-\frac{1}{\tau_t}}\\
&\sim& \frac{M_0}{\beta\omega \tau_\star^\mu (1+\tau_\star^\mu\tau_t^{-\mu})}
\frac{1}{\displaystyle s^\mu-\frac{\tau_\star^\mu\tau_t^{-\mu}
+(\tau_t \beta\omega)^{-1}}{\tau_\star^\mu+\tau_\star^{2\mu}\tau_t^{-\mu}}}
\end{eqnarray}
and thus after Laplace inversion \cite[Eq. (4.10.1)]{gorenflo2014mittag},
\textcolor{black}{
\begin{eqnarray}
M_m(t)&\sim&
\frac{M_0}{\beta\omega \tau_\star^\mu (1+\tau_\star^\mu\tau_t^{-\mu})}
t^{\mu-1}E_{\mu,\mu}(\lambda t^\mu)
\end{eqnarray}}
with $\lambda=[\tau_\star^\mu\tau_t^{-\mu}+(\tau_t \beta\omega)^{-1}]/[\tau_\star^\mu
+\tau_\star^{2\mu}\tau_t^{-\mu}]$.
\textcolor{black}{Since we have $\tau_t\gg \tau_\star$, $\lambda$ simplifies to
$\lambda=1/(\tau_\star^\mu \tau_t \beta\omega)$.} Using
$\mathscr{L}^{-1}\{f(s+1/\tau_t)\}=\exp(-t/ \tau_t)f(t)$, we
find\textcolor{black}{
\begin{equation}
M_m(t)\sim\frac{M_0}{\beta\omega \tau_\star^\mu (1+\tau_\star^\mu\tau_t^{-\mu})}
t^{\mu-1}E_{\mu,\mu}(\lambda t^\mu),~{\tau_\star\ll t\ll \tau_t}.
\end{equation}}

Another class of modification to the models considered above arises for the case
of composite systems, in which two distinct immobile zones with different trapping
time PDFs, $\gamma_1(\tau)$ and $\gamma_2(\tau)$, exist. Analogously to
(\ref{eqCrate}), these systems are described by
\begin{subequations}
\begin{eqnarray}
\frac{\theta_m}{\theta_{im}}\frac{\partial}{\partial t}C_m(x,t)&=&-\omega C_m(x,t)
+\int_0^t[b\gamma_1(t-\tau)+(1-b)\gamma_2(t-\tau)]\omega C_m(x,t)d\tau\nonumber\\
&&+\frac{\theta_m}{\theta_{im}}L(x)C_m(x,t),\\
\frac{\partial}{\partial t}C_{im,1}(x,t)&=&b\omega C_m(x,t)-\int_0^tb\gamma_1(t-\tau)
\omega C_m(x,t)d\tau,\\
\frac{\partial}{\partial t}C_{im,2}(x,t)&=&(1-b)\omega C_m(x,t)-\int_0^t (1-b)
\gamma_2(t-\tau)\omega C_m(x,t)d\tau.
\end{eqnarray}
\label{eqCrateTwoGamams}
\end{subequations}
Here the particle immobilizes into the first immobile zone with probability $b$
and into the second zone otherwise. The combination of two remobilization
processes arises, for instance, in intra-granular diffusion processes, where
mesopores and micropores are present and the latter lead to slow diffusion with
gamma distributed diffusion rates \cite{cunningham1997effects}. We are mainly
interested in the mobile zone, consequently we define
\begin{equation}
\gamma(\tau)=b\gamma_1(\tau)+(1-b)\gamma_2(\tau)
\label{eqTwoGammas}
\end{equation}
and consider (\ref{eqCratea}) only. All observables can be obtained by plugging
the corresponding (composite) trapping time PDFs into the general expressions
that we presented in Section \ref{chGeneralExpressions} and numerically
calculating the Laplace inversion, see the explicit results
in Section \ref{chExpMass}. We note that we calculate all Laplace inversions
using the implementation of the De Hoog method \cite{dehoog1982improved} using
the Python package mpmath \cite{mpmath}.

\section{Connection to fractional models}
\label{chschumer}

We now proceed to show that the EMIM formalism developed here is consistent with
the bi-fractional diffusion equation model \cite{sandev2015distributed} and the 
fractal MIM presented in \cite{schumer2003fractal} in the limit $t\gg\tau_\star$.
The relations between these models and the EMIM are outlined in
Fig.~\ref{figRelationModels}.

\subsection{Connection to bi-fractional diffusion equation and fractal MIM}

To this end we recall our definition of the cumulative function of the waiting
time PDF, $\Psi(t)=\int_t^\infty\gamma(\tau)d\tau$, i.e., the survival probability
in the trapped state. Since $\int_0^\infty\gamma(\tau)d\tau=1$, we have $1-\Psi(t)
=\int_0^t\gamma(\tau)d\tau$. From here we obtain $\gamma(\tau)=-\partial\Psi(t)/
\partial t$ and $\Psi(s)=[1-\gamma(s)]/s$. Now, we aim at rewriting the dynamic
equations (\ref{eqCrate}) of the EMIM in terms of this survival probability. We
start with relation (\ref{eqCratea}) and use integration by parts in the second
term of the right hand side,
\begin{equation}
\int_0^t\gamma(t-\tau)C_m(x,t)d\tau=\Psi(t)C_m(0)+\Psi(0)C_m(t)-\int_0^td\tau
\Psi(t-\tau)\frac{d}{d\tau}C_m(\tau).
\end{equation}
Thus, our model (\ref{eqCratea}) is equivalent to
\begin{equation}
\frac{\partial}{\partial t}C_m(t)+\beta\omega\int_0^t\Psi(t-\tau)\frac{\partial
C_m(\tau)}{\partial \tau}d\tau=-\beta\omega\Psi(t)C_m(0)+L(x)C_m(x,t).
\label{eqOurModelSurvival}
\end{equation}
Now, for our ML model $\gamma(s)=1/(1+(\tau_\star s)^\mu)$ the survival probability
in the Laplace domain reads
\begin{equation}
\Psi(s)=\frac{s^{\mu-1}\tau_\star^\mu}{1+(\tau_\star s)^\mu}.
\label{eqOurSurvival}
\end{equation}
Thus, in our approach $\Psi(t)=E_\mu[-(t/\tau_\star)^\mu]\to e^{-t/\tau_\star}$
for $\mu=1$. The ML function converges to unity when $t\to0$ and decays as the
power-law $t^{-\mu}\tau_\star^\mu/\Gamma(1-\mu)$ at large $t$. If we only retain
the long time asymptotes we
arrive at the model in \cite{schumer2003fractal} in terms of the fractional
Caputo derivative of order $0<\mu<1$ \cite{podlubny}.

For the specific choice $\Psi(t)=\omega e^{-\omega t}$ the fractal model in
\cite{schumer2003fractal} leads to the classical mass transfer model
(\ref{eqCoatsa}). Note that this choice is equivalent to our exponential model
with $\gamma(t)=\omega e^{-\omega t}$. It leads to the linear retardation factor
$(1+\beta)$ \cite{schumer2003fractal} and the dynamic equation
\begin{equation}
(1+\beta)\frac{\partial C_{\mathrm{tot}}}{\partial t}=L(x)C_{\mathrm{tot}},\,\,\,
C_{\mathrm{tot}}(x,0)=\theta_m C_{m,0}(x)
\label{eqSchumerRetardation}
\end{equation}
for the total concentration that we also found in the long-time limit of our
exponential model. In this sense our approach is fully consistent with the fractal
MIM developed in \cite{schumer2003fractal}. However in our EMIM formulation the
trapping time distribution $\gamma(t)$ is a proper PDF including the case of PDFs
with diverging mean; in particular, no divergence at $\gamma(0)$ occurs.

We proceed to analyze the connection of the EMIM to the bi-fractional diffusion
equation. In the long-time limit we can rewrite the total concentration 
(\ref{eqrhoks}) using a ML PDF and the approximation $\gamma(s)=1-\tau_\star^\mu
s^\mu$ to obtain
\begin{equation}
sC_\mathrm{tot}(k,s)-M_0+\beta\omega\tau_\star^\mu s^\mu C_\mathrm{tot}(k,s)-M_0
\beta\omega\tau_\star^\mu s^{\mu-1}=(ikv-k^2D)C_\mathrm{tot}(k,s).
\label{eqctots}
\end{equation}
We can now identify a first-order derivative and a Caputo fractional derivative,
yielding in time-space domain
\begin{equation}
\frac{\partial C_\mathrm{tot}}{\partial t}+\beta_s'\frac{\partial^\mu C_\mathrm{
tot}}{\partial t^\mu}=L(x)C_\mathrm{tot},
\label{eqbifrac}
\end{equation}
which is a bi-fractional diffusion equation, as discussed in 
\cite{sandev2015distributed,maryshev2009non,chechkin2002retarding} and reported 
in \cite{schumer2003fractal} for $\theta_m C_{m,0}(x)=C_\mathrm{tot}(x,0)$,
with a generalized transport operator $L(x)$.

\subsection{Analytical forms of the transport moments}

For small Laplace variable $s$ the Laplace transform of the ML PDF behaves like
$\gamma(s)\sim1-\tau_\star^\mu s^\mu$. Plugging this limiting form into the
Fourier-Laplace transform of the mobile concentration (\ref{eqCms}) we find
\begin{equation}
C_m(k,s)=\frac{M_0}{s+\beta\omega\tau_\star^\mu s^{\mu}-ivk+k^2 D}.
\label{eqCmsSBMB}
\end{equation}
We call this asymptotic form the "fractal model", which coincides with the model
analyzed in \cite{schumer2003fractal} and \cite{sandev2015distributed,maryshev2009non},
as discussed above. We note that even though our model includes the fractal model
in the limit $t\gg\tau_\star$, the bi-fractional models are full models valid for
all $t$ on their own. Therefore, we calculate the mobile mass and the moment for
all $t$ and not only in the limit $t\to\infty$. The advantage of the ML model is
that the trapping PDF (\ref{eqMLt}) is well defined in the limit $t\to0$.

To find the mobile mass using the fractal model, we set $k=0$ in $C_m(k,s)$ in
Eq.~(\ref{eqCmsSBMB}) and use the properties of the ML function
\cite[Eq. (3.7.8)]{gorenflo2014mittag}, yielding
\begin{equation}
M_m(t)=M_0E_{1-\mu}(-\beta\omega\tau_\star^\mu t^{(1-\mu)}).
\label{eqSBMBmass}
\end{equation}
For the unnormalized first moment we use the fractal model and the general formula
(\ref{eqxscaled}) for $\langle x_m(s)\rangle_u$ to find
\begin{equation}
\langle x_m(s)\rangle_u=\frac{v}{(s+\beta\omega(\tau_\star s)^{\mu})^2}=\frac{vs^{
-2\mu}}{(s^{1-\mu}+\beta\omega\tau_\star^\mu)^2},
\label{eqxscaledL2}
\end{equation}
which we transform to time domain using \cite[Eq. (2.5)]{prabhakar1971singular} 
\begin{equation}
\langle x_m(t)\rangle_u=vtE^2_{1-\mu,2}(-\beta\omega\tau_\star^\mu t^{(1-\mu)}).
\label{eqSBMBx_un}
\end{equation}
Dividing the unnormalized first moment (\ref{eqSBMBx_un}) by the mobile mass 
(\ref{eqSBMBmass}) normalizes the first moment,
\begin{equation}
\langle x_m(t)\rangle=\frac{vtE^2_{1-\mu,2}(-\beta\omega\tau_\star^\mu t^{(1-\mu)})}
{E_{1-\mu}(-\omega\beta\tau_\star^\mu t^{(1-\mu)})}.
\label{eqSBMBxnormalized}
\end{equation}
We plug the fractal model with $\gamma(s)\sim 1-(\tau_\star s)^\mu$ into the general
relation (\ref{eqximuGeneral}) between $\langle x^n_m\rangle_u$ and $\langle x^n
_{im}\rangle_u$ for $n=1$, obtaining
\begin{equation}
\langle x_{im}\rangle_u=\omega\beta v\frac{1}{(s+\beta\omega(1-\gamma(s)))^2}
\frac{1-\gamma(s)}{s}=\omega\beta v\frac{\tau_\star^\mu s^{\mu-1}}{(s+\beta
\omega\tau_\star^\mu s^\mu)^2}.
\label{eqximu}
\end{equation}
In time-domain we find using (\ref{eqximu}) and \cite[Eq. (2.5)]{prabhakar1971singular} 
that
\begin{equation}
\langle x_{im}\rangle_u=\omega\beta\tau_\star^\mu t^{2-\mu}E_{1-\mu,3-\mu}^2
\left(-\beta\omega\tau_\star^\mu t^{1-\mu}\right).
\label{eqximt}
\end{equation}
Dividing by the immobile fraction mass $M_{im}(t)/M_0=1-M_m(t)/M_0$ yields the
normalized first moment,
\begin{equation}
\langle x_{im}\rangle=\frac{\omega\beta\tau_\star^\mu t^{2-\mu}E_{1-\mu,3-\mu}^2
\left(-\beta\omega\tau_\star^\mu t^{1-\mu}\right)}{1-E_{1-\mu}(-\beta\omega\tau_
\star^\mu t^{(1-\mu)})}.
\label{eqximtnormalized}
\end{equation}

Consider next the unnormalized second moment obtained via the second derivative
of $C_m(k,s)$ in Eq.~(\ref{eqCmsSBMB}),
\begin{equation}
\langle x_m^2(s)\rangle_u=\frac{2D }{(s+\beta\omega\tau_\star^\mu s^{\mu})^2}
+\frac{2v^2}{(s+\beta\omega\tau_\star^\mu s^{\mu})^3}=\frac{2Ds^{-2\mu}}{(s^{
1-\mu}+\beta\omega\tau_\star^\mu)^2}+\frac{2v^2s^{-3\mu}}{(s^{1-\mu}+\beta\omega
\tau_\star^\mu)^3},
\label{eqx2scaledL2}
\end{equation}
which we transform back to the time domain using
\cite[Eq.~(2.5)]{prabhakar1971singular},
\begin{equation}
\langle x_m^2(t)\rangle_u=2Dt E^2_{1-\mu,2}(-\beta\omega\tau_\star^\mu t^{(1-\mu)})
+2t^2v^2E^3_{1-\mu,3}(-\beta\omega\tau_\star^\mu t^{(1-\mu)}).
\label{eqSBMBxxUnnormalizedwV}
\end{equation}
With the mobile mass (\ref{eqSBMBmass}) we normalize (\ref{eqSBMBxxUnnormalizedwV})
to
\begin{equation}
\langle x_m^2(t)\rangle=2Dt\frac{E^2_{1-\mu,2}(-\beta\omega\tau_\star^\mu t^{(1
-\mu)})}{E_{1-\mu}(-\beta\omega\tau_\star^\mu t^{(1-\mu)})}+2t^2v^2\frac{E^3_{1
-\mu,3}(-\beta\omega\tau_\star^\mu t^{(1-\mu)})}{E_{1-\mu}(-\beta\omega\tau_\star
^\mu t^{(1-\mu)})}.
\label{eqSBMBxxnormalizedwV}
\end{equation}

We find the second central moment of the fractal model by using relations
(\ref{eqSBMBxnormalized}) and (\ref{eqSBMBxxnormalizedwV}),
\begin{eqnarray}
\langle (x_m(t)-\langle x_m(t)\rangle)^2\rangle&=&2Dt\frac{E^2_{1-\mu,2}(-\beta
\omega\tau_\star^\mu t^{(1-\mu)})}{E_{1-\mu}(-\beta\omega\tau_\star^\mu t^{(1
-\mu)})}+2v^2t^2\frac{E^3_{1-\mu,3}(-\beta \omega\tau_\star^\mu t^{(1-\mu)})}{E_{
1-\mu}(-\beta\omega\tau_\star^\mu t^{(1-\mu)})}\nonumber\\
&-&v^2t^2\left(\frac{E^2_{1-\mu,2}(-\beta\omega\tau_\star^\mu t^{(1-\mu)}}{E_{
1-\mu}(-\beta\omega\tau_\star^\mu t^{(1-\mu)})})\right)^2.
\end{eqnarray}
We note that the asymptotics of the moments presented in this chapter can be obtained 
by rewriting the three-parametric ML functions in terms of two-parametric ML
functions with \cite[Eq. (2.4)]{prabhakar1971singular} and approximating them up to
second order with \cite[Eq. (6.11)]{haubold2011mittag}. These limits match our
results for the EMIM ML model in Section \ref{chML} in Eqs.~(\ref{eqMassLimit}),
(\ref{eqxmlimit}), and (\ref{eqSBMBMSDlimit2}) and what has been reported previously
\cite{zhang2008moments}.

\section{Comparison to experiments}
\label{chExperiments}

We apply our model to two different experimental data sets that we discuss in detail.
The study reported in \cite{goeppert2020experimental} probes fluorescent dye in an
alpine karst aquifer in the Hochifen-Gottesacker area (Austria). The dye was injected
into actively sinking surface water and measured at two karst springs up to 7400 m
far away \cite{goeppert2020experimental}. We show the resulting BTC of the data set
IP Gb (2) in Fig.~\ref{figBTCfit}. It was measured 3500 m downstream from the
injection point and was previously reported in \cite{goldscheider2005fold,
goeppert2020experimental}. Our aim is to obtain the moments of the mobile
concentration from this BTC, because BTCs are commonly measured while moments
provide important additional information on the transport dynamics
\cite{goeppert2020experimental,edery2014origins,aubeneau2016biofilm,
berkowitz2006modelling,gouze2008non-fickian}.

One crucial idea of our model is the division into mobile and immobile particles.
However, the tracers are not detected whilst moving through the karst aquifer and
hence we cannot directly compare our predicted mobile mass decay to experiments
that only measure BTCs. Therefore we consider a second experiment where the
tracer concentration profile is measured. This experiment is the MADE-1 (first
macrodispersion experiment) \cite{boggs1992field,adams1992field}. The authors of
this study realized a 48 h pulse injection of bromide into a heterogeneous aquifer
near Columbus, Mississippi (USA). A network of multilevel sampling wells covering
around 300 m along the flow direction with approximately 6000 sampling points
allowed the observation of the plume profile at eight snapshots up to 594 days
after injection. Using linear interpolation between sampling points the authors
obtained the total measured mass by integrating over all three spatial dimensions.
The total recovered mass exceeds the initial input mass, which the authors explain
by a "spurious hydraulic connection among [the sampling stations]" or higher
concentrations in regions with higher hydraulic conductivity and subsequent
inaccurate linear interpolation \cite{adams1992field}. Nevertheless, a power-law
tail was clearly observed in the decay of recovered mass, as demonstrated in
\cite{schumer2003fractal}. In addition to the mobile mass, the authors obtain
the moments of the tracer distribution in the MADE-1 \cite{adams1992field}.
Notably, the plume consists of a virtually stationary distribution with a slowly 
decaying shoulder \cite{adams1992field}.

\subsection{Mobile mass}
\label{chExpMass}

\begin{figure}
\includegraphics[width=12cm]{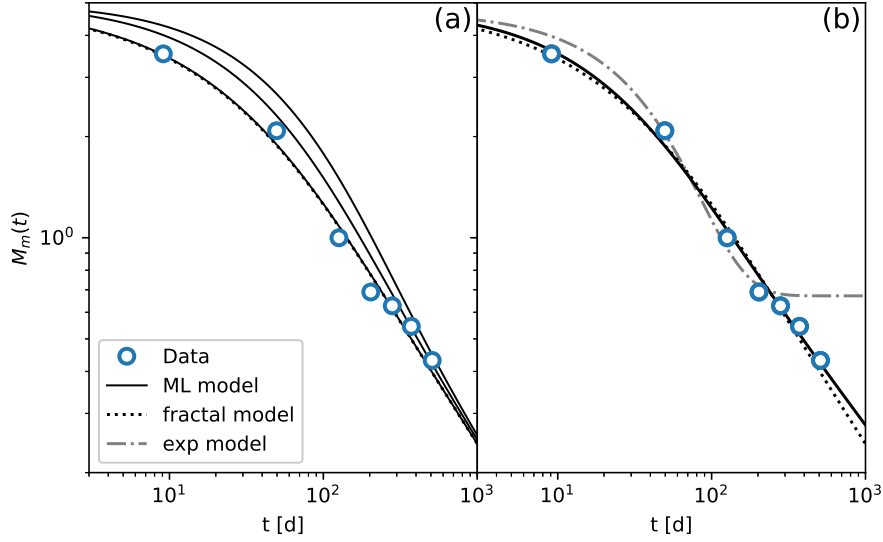}
\caption{Fraction of measured mass in MADE-1 experiment. Data points are taken
from \cite{schumer2003fractal}, extracting the data from the PDF paper file.
\textbf{(a)} The fractal MIM is fitted to the data, the parameters are $\beta_s
=0.08\mathrm{days}^{-0.67}$ and $\mu=0.33$ with initial mobile mass of $5$, as
as obtained in \cite{schumer2003fractal}. For the ML model (\ref{eqMLt}) we
choose $\mu=0.33$ and $\beta\omega\tau_\star^\mu=0.08\mathrm{day}^{\mu-1}$
such that both the ML and fractal models yield the same long time asymptotic
behaviors. The asymptote does not depend on the ratio $\beta\omega/\tau_\star
^\mu$, for which we choose $1/10$, $1/5$ and $10$ from top to bottom. This ratio
affects the short time behavior $\simeq\exp(-\omega\beta t)$. Small values of
$\tau_\star$ lead to an early asymptotic behavior and hence a match with the
fractal model appears sooner. \textbf{(b)} The exponential model, ML model, and
fractal MIM are fitted individually to the data. The asymptotic power-law of the
ML EMIM model has a different exponent of $\mu-1=0.42-1=-0.58$ compared to the
fractal model with $\mu-1=0.33-1=-0.67$. The exponential model has the parameters
$\beta=6$, $M_0=4.7$, and $\omega=0.003/\mathrm{day}$, as obtained in 
\cite{schumer2003fractal}. The associated parameters of the remaining models are
listed in Tab.~\ref{tabPars}.}
\label{figmasscompare}
\end{figure}

\begin{table}
\begin{tabular}{|l|l|l|l|l|l|l|l|l|l|}
\hline
Model & $M_0$ & $\mu$ & $\beta\omega$ [/day] & $\tau_\star$ [days]& $\tau_t$ [days]&
$\omega_1$ [/day]& $\omega_2$ [/day]& $b$ & $R^2$ \\\hline
ML & 4.833 & 0.417 & 0.060 & 9.549 & & & &  &0.992\\
truncated & 4.726 & 0.226 & 0.913 & 0.934  & 11530 & & &  &0.994\\
ML+exp& 4.2 & 0.104 &0.0227    & 25.18 & &0.0052 & & 0.56 & 0.998\\
exp+exp& 3.78 &  &  0.01983    &  & & 0.00182 &  5713 & 0.61 & 0.989\\\hline
\end{tabular}
\caption{Fit parameters and coefficient of determination of fits to $M_m(t)$ from
Figs.~\ref{figmasscompare}(b) and \ref{figModifiedGamma}. The parameters $\beta$
and $\omega$ only appear as a product in the mobile concentration and hence cannot
be determined separately.}
\label{tabPars}
\end{table}

Fig.~\ref{figmasscompare} shows the measured mobile mass decay of the MADE-1
\cite{adams1992field} as circles. In panel (a) we show the best fit of the fractal
MIM from \cite{schumer2003fractal} (see section \ref{chschumer}) along with our ML
model. For the latter we did not fit the data but choose the model parameters such
that the ML model has the same asymptotic long time behavior as the fractal model.
This is achieved for the parameters $\mu=0.33$, $\beta\omega\tau_\star^\mu=0.08
\mathrm{day}^{\mu-1}$ and $M_0=5$ for the mobile mass. For the ML model we show the
numerical Laplace inversion of the mobile mass in Laplace space (\ref{eqMs}) using
$\gamma_{ML}$. At short times all models are dominated by the identical initial
value $M_0$, hence the ML and fractal models differ only at intermediate time scales
of around 50 days. Note that as long as the product $\beta\omega\tau_\star^\mu$
remains constant, the same long-time limit is reached. Therefore, we can choose
different ratios $\beta\omega/\tau_\star^\mu$. From top to bottom we use in
Fig.~\ref{figmasscompare} the values $1/10$, $1/5$ and $10$. A small ratio will
decrease the initial decay $\simeq \exp(-\beta\omega t)$, while a large ratio
corresponding to small $\tau_\star$ leads to earlier appearance of the asymptotic
behavior, and for the ratio $10$ the ML model coincides with the fractal model.

In Fig.~\ref{figmasscompare}(b) we show a fit with our model (\ref{eqClassicMass})
with an exponential trapping time distribution with $\beta=6$, $M_0=4.7$ and $\omega=
0.00310811/\mathrm{day}$. These parameters correspond to a fit to the data shown in
\cite{schumer2003fractal}, where a model matching (\ref{eqCandS}) was used. The
fit does not describe the data well, because it reaches the steady state value 
(\ref{eqClassicLargeTmass}), in contrast to the continued decay shown by the data.
In addition, we show fits of both the fractal and ML model to the MADE-1 data. Both
models describe the data well, as demonstrated by the coefficient of determination
$R^2=0.992$ (we calculate all coefficients of determination using the Python module
\texttt{scikit learn} \cite{sklean2021grisel}). In Tab.~\ref{tabPars} we show the
fit parameters, observing no significant difference in goodness of fit between the
ML and fractal model. We note, however, that $\mu$ differs: it is $0.33$ for the
fractal model and $0.42$ for the ML model. This observation demonstrates that the
fully quantitative behavior of the seemingly very similar models is indeed
notably different.

\begin{figure}
\includegraphics[width=12cm]{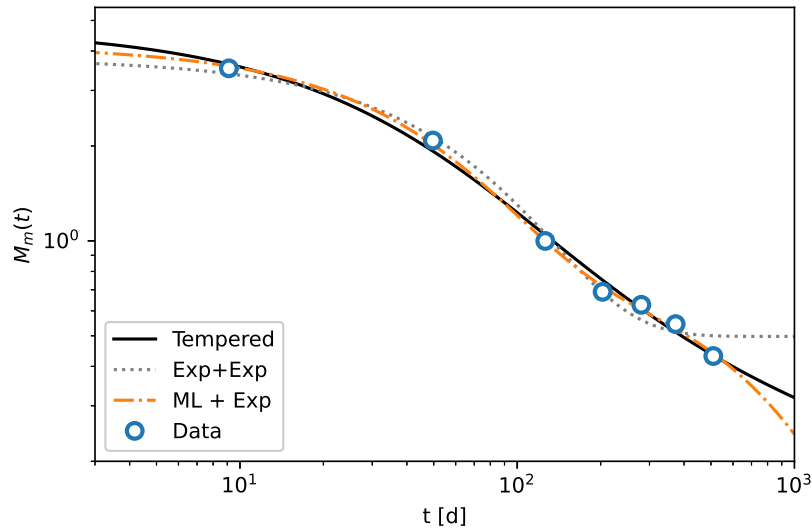}
\caption{Fitted extended models. The tempered ML model $\gamma^*_{ML}(s)=[1+(
\tau_\star/\tau_t)^\mu]/(1+(\tau_\star s+\tau_\star/\tau_t)^\mu)$ (solid line) and two
combinations of $\gamma$ are fitted to the experimental mobile mass decay of the
MADE-1 experiment \cite{adams1992field}. At around 900 days, the deviation from
the power-law trend is visible in the tempered model. The combination of two
exponential functions (dotted line) behaves quite closely to the pure exponential
model in Fig.~\ref{figmasscompare}. Combining the ML and exponential models yields
the dash-dotted orange line. It is the only considered model capable of describing
the shoulder of the data at around 300 days. Fit parameters are given in
Tab.~\ref{tabPars}.}
\label{figModifiedGamma}
\end{figure}

Fig.~\ref{figModifiedGamma} shows fits using our extended models from Section
\ref{chextensions} to the MADE-1 data \cite{schumer2003fractal,adams1992field},
see Tab.~\ref{tabPars} for the fit parameters. First we consider the composite
model with two exponential terms, $\gamma(\tau)=b\omega_1\exp(-\omega_1\tau)+
(1-b)\omega_2\exp(-\omega_2 \tau)$ with $0<b<1$. The result is shown by the
dotted line, which quantitatively behaves quite similarly to the exponential
model. It approximates all but the last data point well with a coefficient of
determination of $R^2=0.988$, which is notably worse than all models containing
power-law waiting times, see Tab.~\ref{tabPars}. This indicates the necessity of
including long-tailed trapping time PDFs for this data. Of course, adding
additional exponentials would improve the fit, however, on the cost of a
larger number of fit parameters.

The second composite form that we consider reads in Laplace space 
\begin{equation}
\gamma(s)=\frac{b}{1+(\tau_\star s)^\mu}+(1-b)\frac{\omega_1}{\omega_1+s}
\label{eqMLandExp}
\end{equation}
and corresponds to the combination of an exponential and an ML trapping time PDF.
In Fig.~\ref{figModifiedGamma} the dash-dotted orange line shows the best fit
using this model, with coefficient of determination of $R^2=0.998$. In fact this
is the only model considered here capable of reproducing the apparent shoulder
in the data around 200 days. Concurrently, the long time behavior exhibits a
scaling exponent $\mu$ that is significantly different from the pure ML model.
We highlight that both the truncated ML model (\ref{eqgammat}) with $R^2=0.993$
and the combination of the ML and exponential model (\ref{eqMLandExp}) with $R^2
=0.998$ fit the data better than the ML model alone ($R^2=0.992$). This points at
the fact that the data indeed encode finite size effects needing a tempering of
the power-law tail of the trapping time PDF. However, we stress that we fit to
seven data points only and the extended models have more parameters than the ML
or exponential model. Therefore, the extended models might be subject to overfitting.
Improved data will be needed to be more accurate in this interpretation.

\subsection{Breakthrough curves (BTCs)}

\begin{figure}
\includegraphics[width=12cm]{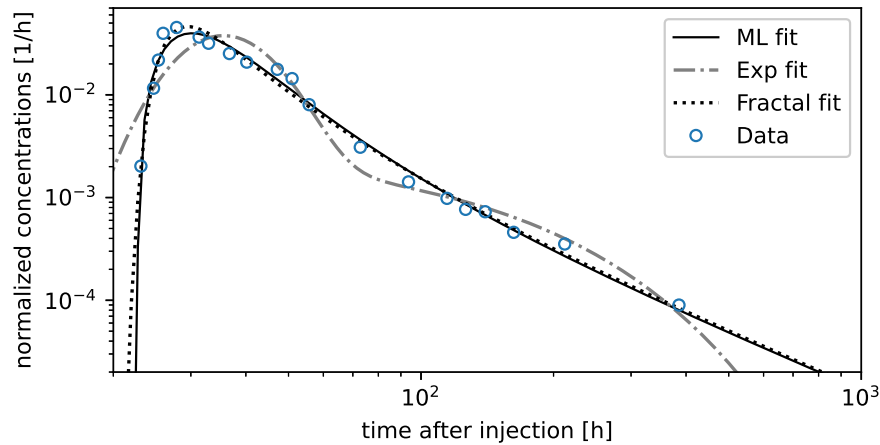}
\caption{Fits to the BTC for fluorescent dye measured 3500 m downstream from the
injection point in the Schwarzwasser experiment \cite{goldscheider2005fold,
goeppert2020experimental}. The solid line shows a fit using the ML model with
$R^2=0.990$, the dashed line a fit using the fractal model with $R^2=0.990$. Both
differ mainly in the peak height. An exponential trapping time PDF yields a
significantly worse fit ($R^2=0.940$) with an additional shoulder and an
exponential decay instead of the power-law decay at long times. Data taken from
\cite[IP Gb (2)]{goeppert2020experimental}. Fit parameters are listed in
Tab.~\ref{tabParsBTC}.}
\label{figBTCfit}
\end{figure}

In Fig.~\ref{figBTCfit} we show the BTC of the IP Gb (2) experiment, in which
fluorescent dye travels in the underground aquifer in the Schwarzwasser valley
\cite{goeppert2020experimental}. All fit parameters are listed in
Tab.~\ref{tabParsBTC}. A fit using the ML trapping time PDF describes the data well
with $R^2=0.990$. The fractal model describes the data equally well with the same
$R^2=0.990$. Its peak is slightly higher than that of the ML model. The classical
model fails to describe the power-law decay at long times and hence yields the
notably worse value $R^2=0.940$, with an unlikely high fit value for the diffusion
coefficient, $D_\mathrm{exp}=9527\mathrm{m}^2/$h, and a considerably lower $v_
\mathrm{exp}=$94.9 m/h. In this model, the tracers diffuse very fast while the
advection is slow. This way, the exponential fit
compensates for the lack of the power-law decay. Note also the exponential decay
in the fit of the exponential model and the fact that it completely misses the
short time behavior. We conclude that power-law tails introducing a wide range
of time scales appear necessary for a proper description of the experimental data.
We use the parameters from the fits to the BTC for the remainder of this work.

\begin{table}
\begin{tabular}{|l|l|l|l|l|l|l|l|l|}
\hline
Model & $M_0$ & $\mu$ & $\beta\omega$ [/h] & $\omega$ [/h] & $\tau_\star$ [h] &
$v$ [m/h]& $D$ [$\mathrm{m}^2/$h]& $R^2$\\\hline
ML & 155.7 & 0.737 &  0.406  &  & 0.922 & 151.4 & 35.2 & 0.990 \\
limit & 242.8 & 0.771 & 0.841   & & 1.00 & 232.9& 21.8 & 0.990 \\
exp & 96.3 & & .006 & 0.011 & & 94.9& 9527.0 & 0.940 \\
\hline
\end{tabular}
\caption{Fit parameters and coefficients of determination for the BTCs shown in
Fig.~\ref{figBTCfit}. The parameters $\beta$ and $\omega$ only appear as a product
in the mobile concentration of the ML and fractal model and hence cannot be
determined separately.}
\label{tabParsBTC}
\end{table}

\subsection{Concentration profile}

\begin{figure}
\centering
\includegraphics{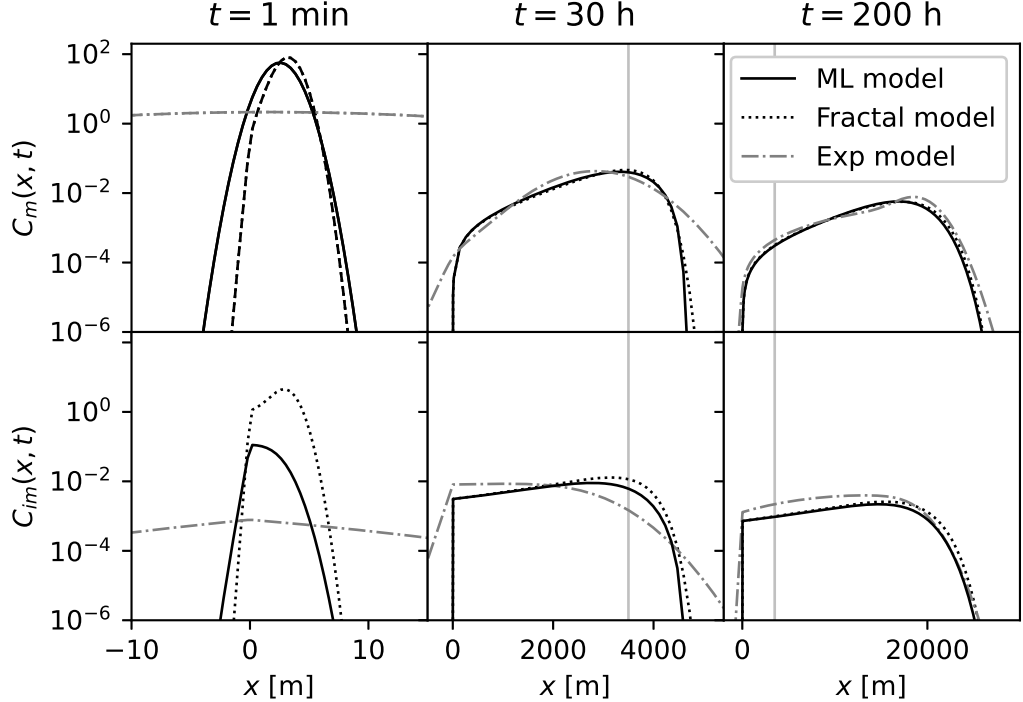}
\caption{Logarithmic representation of the concentration profiles of the mobile
and immobile particles 1 min, 30 h, and 200 h after injection from left to right
panels, respectively. Note the different ranges of the horizontal axes for each
delay time. At short times all models yield a Gaussian mobile distribution, which
is increasingly skewed at later times. The diffusion coefficient  $D_\mathrm{exp}
=9527\mathrm{m}^2/$h of the exponential model is exceptionally high compared to
the ML model ($D_\mathrm{ML}=35.2\mathrm{m}^2/$h) and the fractal model ($D_
\mathrm{limit}=21.8\mathrm{m}^2/$h). Thus the concentration of the exponential
model has spread significantly further than the other models at shorter times.
The immobile particle fraction has a non-Gaussian distribution at all times and
has cusps at $x=0$, which are typical transport features for systems with
diverging waiting times \cite{harvey,report,sandev2015distributed}.
The parameters of all models correspond to the fit to the BTC in
Fig.~\ref{figBTCfit}. The BTC was measured at 3500 m, as marked by the vertical line.
See Tab.~\ref{tabParsBTC} for fit values.}
\label{figConcentrations}
\end{figure}

Next, we focus on the spatial distribution of the solute, i.e., the plume profiles.
In Fig.~\ref{figConcentrations} we show the concentrations corresponding to the
BTC fits in Fig.~\ref{figBTCfit}. We logarithmically present the Laplace inversion
of the mobile concentration (\ref{eqCmxs}) in the upper row for the ML model (solid
line), the fractal model (dotted line), and the exponential model (dash-dotted line).
The first column shows the short-time behavior 1 min after injection. The ML and
fractal models show a comparatively narrow bell-shaped behavior while the exponential
model already exhibits a considerable spread due to its high value of $D$. At
intermediate times of 30 h (middle column) all mobile concentrations are
increasingly skewed to the right and are significantly non-Gaussian (note the
different ranges of the horizontal axes for different $t$). The vertical line
denotes the position at which the experimental BTC was measured (3500 m downstream
from the injection point). Quick decays to zero concentration around 0 m and 4000 m
characterize the skewed limit and ML model concentrations. At 200 h, the difference
between the three models is quite small. At the measurement position, the limit and
ML model are particularly similar.

The immobile concentrations in the second row of Fig.~\ref{figConcentrations} are
pronouncedly non-Gaussian at all times. We obtain them by taking a numerical
Laplace inversion of the mobile concentration (\ref{eqCmxs}) plugged into the
general relation (\ref{eqCimks}) between the mobile and immobile concentration.
At $t=$1 min the fractal model's immobile concentration is one order of magnitude
higher than that of the ML model. In addition, the peak of the ML model is close
to $x=$0 m while the peak of the fractal model is around $x=$5 m. At 30 h the ML
and fractal models are very similar, have a sharp rise from zero to approximately
$10^{-3}$ at 0 m and show a peak around 3000 m. In contrast, the exponential model
has its peak close to 0 m and falls of monotonically in both directions. At 200 h
the ML and fractal models almost coincide and have qualitatively the same shape as
for 30 h, although they have spread up to 22 km. Notably the exponential model has
a similar concentration.

In the next subsection we characterize these concentrations further in terms of
their first and second moments.

\subsection{Moments}

\begin{figure}
\includegraphics{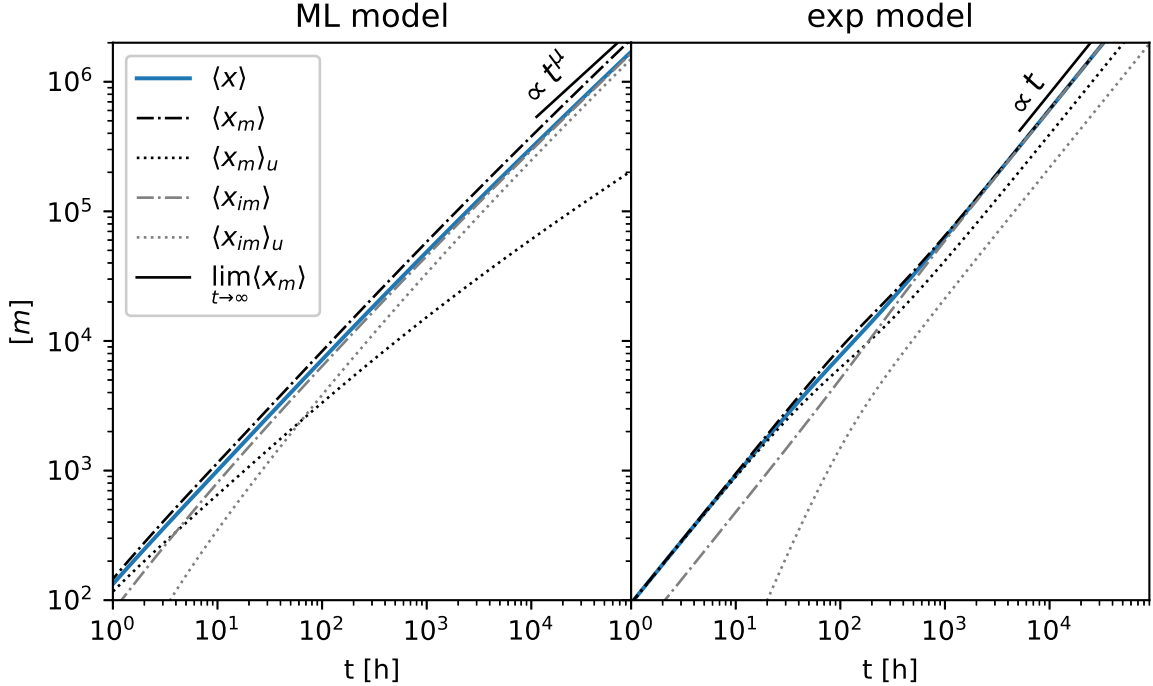}
\caption{Comparison of the normalized ($\langle x_{m,im}\rangle$), unnormalized
($\langle x_{m,im}\rangle_u$) and total first moment $\langle x\rangle$ for the
mobile and immobile tracers on a double logarithmic scale. The left panel shows
the ML model, the right one shows the exponential model. In both panels, both
$\langle x_m\rangle_u$ and $\langle x_m\rangle$ are a good approximation for
$\langle x\rangle$ at short times, due to the high mobile fraction. The same
holds for long times for the immobile moment. In the ML model, the first moment
of the mobile particles has the same power-law behavior $\simeq t^\mu$ as the
immobile tracers but with a larger coefficient. The black solid line shows the
asymptote (\ref{eqxmlimit}) of $\langle x_m\rangle$ as a guide to the eye. The
parameters correspond to the fit to the BTC in Fig.~\ref{figBTCfit} as listed
in Tab.~\ref{tabParsBTC}.}
\label{figxall}
\end{figure}

We show the first moment $\langle x\rangle$ in Fig.~\ref{figxall} with
parameters corresponding to the fit of our model to the BTC in
Fig.~\ref{figBTCfit}. In addition to $\langle x\rangle$ we show the first moment
of the mobile and immobile tracers, $\langle x_m\rangle$ and $\langle
x_{im}\rangle$ respectively. For the exponential model we use the analytic
expressions (\ref{eqClassicx}) for $\langle x_m\rangle_u$ and
(\ref{eqClassicxNorm}) for $\langle x_m\rangle$. All remaining first moments are
obtained through Laplace inversion of the general expressions (\ref{eqxscaled})
for $\langle x_m\rangle_u$ and (\ref{eqximuGeneral}) for $\langle
x_{im}\rangle_u$. Subsequent normalization with the mobile mass (\ref{eqMm})
yields $\langle x_m\rangle$ and $\langle x_{im}\rangle$ for the ML model.

\begin{figure}
\includegraphics{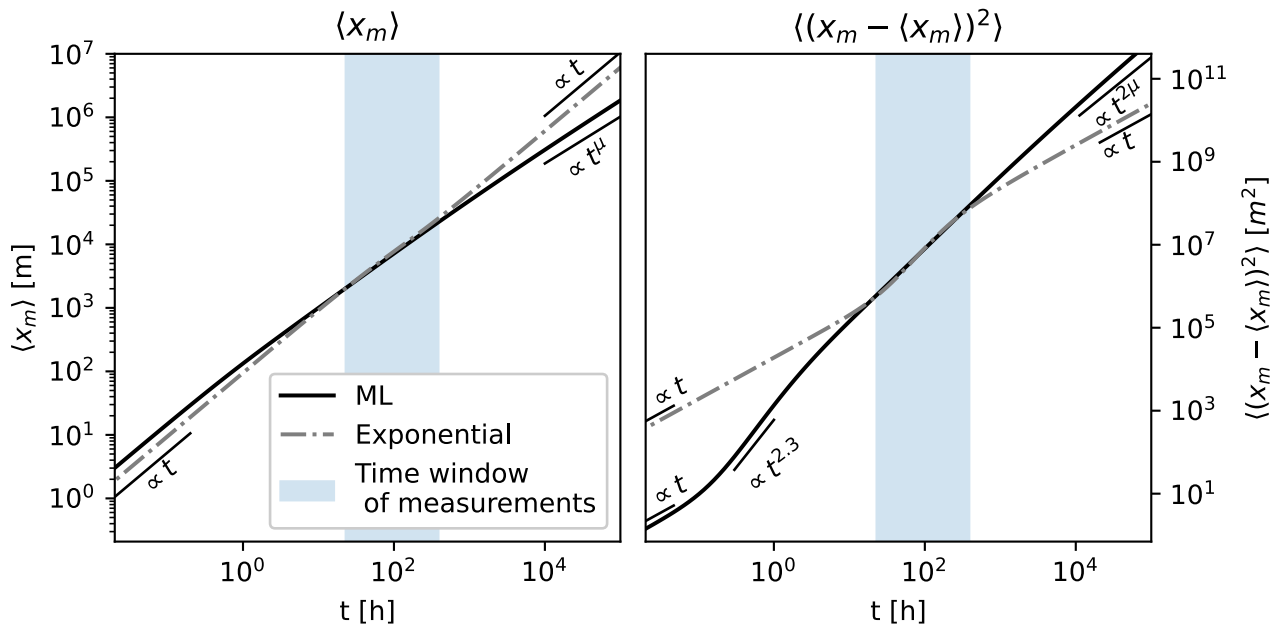}
\caption{First and second central moments. In the left panel we show the first
mobile moment using the exponential and ML model, see Fig.~\ref{figxall} for
details. In the right panel the second central moment of the ML and exponential
models scales like $2Dt$ at
short times. The exponential model is linear in the long-time limit, as well. The
ML model grows $\simeq t^{2\mu}$ for $t>$10 h. The parameters of both models in
both panels are taken from the fit to the BTC in Fig.~\ref{figBTCfit} and are
listed in Tab.~\ref{tabParsBTC}. In the shaded regions corresponding to the time
window in which the data of the fitted BTCs were taken, the exponential model
displays apparent anomalous diffusion and almost coincides with the power-law
from the ML model.}
\label{figallXX_X}
\end{figure}

The first moment $\langle x\rangle$ of the total mass in Fig.~(\ref{figxall}) in
the Appendix demonstrates a crossover from linear to power-law ($\simeq t^\mu$)
scaling in time when using the ML model. At short times it matches $\langle
x_m\rangle$ and $\langle x_m\rangle_u$ while coinciding with $\langle
x_{im}\rangle$ and $\langle x_m \rangle_u$ at long times, as expected by the
immobilization of all tracers in the long-time limit. The moments
\textcolor{black}{$\langle x_m\rangle$ and $\langle x_{im}\rangle$} have the same
long-time power-law behavior, albeit with different coefficients. The solid
black line in the left panel of Fig.~\ref{figxall} shows the long-time limit of
$\langle x_m\rangle$ using the ML model as given by expression
(\ref{eqxmlimit}). In contrast, the center of mass of the mobile plume of the ML
model is ahead at all times after injection. In Fig.~\ref{figxallAppendix} the
described behavior is easier to discern with a significantly lower $\mu=0.33$.
We use the parameters of the fit to the mobile mass in \cite{schumer2003fractal}
of Fig.~\ref{figmasscompare} along with $v=$0.8 m/day and $D=$4
$\mathrm{m}^2/$day.  The advection speed $v$ was measured in the experiment and
$D$ is our estimate.  In the time interval, in which the BTC data was collected,
the first mobile moments almost coincide and grow non-linearly $\simeq t^{\mu}$,
including the exponential model, as shown in Fig.~\ref{figallXX_X}.

\begin{figure}
\includegraphics{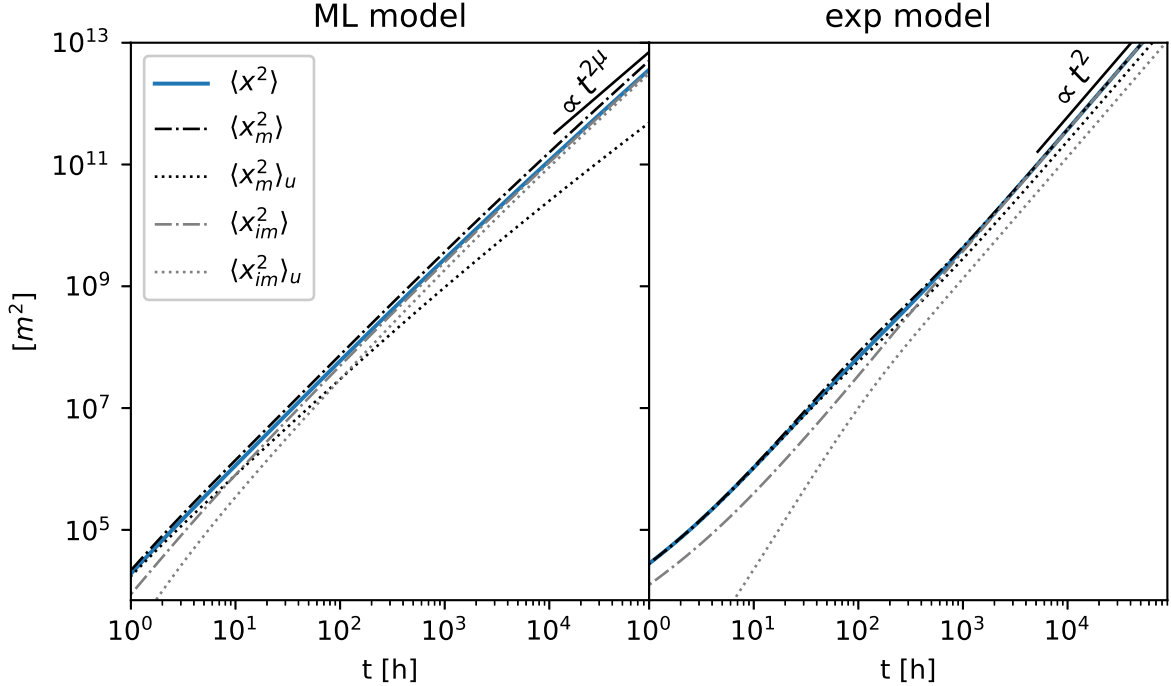}
\caption{Comparison of the normalized ($\langle x^2_{m,im}\rangle$), unnormalized
($\langle x^2_{m,im}\rangle_u$), and total second moment $\langle x^2\rangle$ for
the mobile and immobile tracers on a double logarithmic scale. All models yield
$\langle x^2(t)\rangle=\langle x^2(t)\rangle_u\approx 2Dt$ at short times. The
exponential trapping time PDF leads to the long-time growth $v^2t^2/(1+\beta)^3$
and $v^2t^2/(1+\beta)^2$ of $\langle x^2_m\rangle_u$ and $\langle x^2_m\rangle$,
respectively. At long times the ML model show anomalous diffusion, $\langle x^2
\rangle\simeq t^{2\mu}$, i.e., superdiffusion for $1/2<\mu<1$. The black solid
line shows the asymptote (\ref{eqSBMBMSDlimit2}) of $\langle x_m^2\rangle$ as a
guide to the eye. Parameters are the same as in Fig.~\ref{figmasscompare}(a).}
\label{figallXXcombine}
\end{figure}

In Fig.~\ref{figallXXcombine} we show the second moments corresponding to the
BTC fits from Fig.~\ref{figBTCfit}. We obtain the moments through Laplace
inversion of the general expressions (\ref{eqx2scaled}) for $\langle
x^2_m(s)\rangle_u$ and (\ref{eqximuGeneral}) for $\langle x^2_{im}\rangle_u$
after normalization with $M_m(t)$ or $M_{im}=M_0-M_m$ (obtained from the general
expression (\ref{eqMs}) for $M_m(s)$). The second moment of the total
concentration obtained from the general expression (\ref{eqx}) for $\langle
x^n\rangle$ agrees with $\langle x^2_m\rangle$ both for the ML and exponential
order model at short times as almost all tracers are mobile at this time. In
contrast to the ML model the first order model shows a crossover from linear to
quadratic behavior. Around 1000 h, $\langle x^2\rangle$ and $\langle
x^2_{im}\rangle$ coincide and grow proportionally to $t^{2\mu}$. The second
mobile moment has the same power-law growth, albeit with a higher prefactor, as
demonstrated by the long time limit (\ref{eqSBMBMSDlimit2}) of $\langle x_m^2
\rangle$ shown as a solid black line. In the exponential model, all normalized
second moments grow quadratically and overlap around 1000 days, as expected by
the equivalence of $C_m$ and $C_{im}$ (\ref{eqExpCmCim}) at long times.
Fig.~\ref{figallXX_X} shows the first and second central moment of the mobile
concentration the latter of which is obtained via $\langle(x_m-\langle
x_m\rangle) ^2\rangle=\langle x^2_m\rangle-\langle x_m\rangle^2$. Note that only
the analysis of normalized second central moments is meaningful, therefore we do
not explicitly study the second central moment of the unnormalized moments. At
short times, both the ML (solid line) and exponential model (dash-dotted line)
grow linearly. The prefactor $D$ of the latter is two orders of magnitude larger
and the linear regime lasts until around 10 h. The ML model yields linear growth
up to 6 min and transitions to the power-law $t^{2\mu}$ after a transient growth
proportional to $t^{2.3}$ around 30 min. In the range from 20 h to 300 h we
fitted our models to the BTCs as shown in Fig.~\ref{figBTCfit}. In this range,
that we highlight by the shaded areas in Fig.~\ref{figallXX_X}---and only in
this range---the first and second central moments of the two models almost
coincide. Hence, the exponential model can show transient anomalous diffusion.
This is a remarkable result, that, to the best of our knowledge, has not been
pointed out before. Outside this time window at longer times the moments
demonstrate distinct differences. The exponential model shows normal diffusion
with $\langle(x_m-\langle x_m\rangle)^2\rangle\simeq t$, while the mobile
particles in the ML model spread faster with $\langle(x_m-\langle
x_m\rangle)^2\rangle\simeq t^{2\mu}$.

Here we analyzed anomalous diffusion using the first and second moments, while
higher moments can reveal non-Gaussianity properties of the concentration
profile.  In Appendix \ref{chkurtosis} we calculate the skewness and kurtosis.
These clearly show that after short times both the mobile and immobile
concentration are non-Gaussian when using the ML model. For $t\to\infty$, we
find that the skewness and kurtosis only depend on $\mu$. Notably, both appear
to be discontinuous at $\mu=1$ and jump to their respective Gaussian values.

\section{Conclusion}
\label{secconc}

We introduced and discussed the extended mobile-immobile model (EMIM) for tracer
motion in which the residence time in the immobile domain is drawn from a general
trapping time PDF $\gamma(t)$. The mobile times are chosen to always follow an
exponential distribution. A system with an \emph{exponential trapping time PDF\/}
can then be rewritten in terms of a classical first order mass transfer model
\cite{coats1964deadend}. We consider the initial condition when all particles are
mobile after a pulse injection. This leads to a Gaussian mobile plume at short
times for any $\gamma(t)$. At intermediate times particles in the mobile phase
are trapped in the immobile zone following an exponential trapping time PDF which
renders the mobile concentration non-Gaussian and the moments grow non-linearly.
The second central moment exhibits an apparent anomalous diffusion in this time
regime. In the long-time limit the mobile and immobile concentrations coincide and
we recover normal diffusion with a rescaled time $t/(1+\beta\langle\tau\rangle
\omega)$, where $\beta$ denotes the ratio of immobile to mobile volume, $\langle
\tau\rangle$ represents the average immobilization time, and $\omega$ in time
stands for the mass transfer coefficient.

When using a \emph{scale-free trapping time PDF} with power-law tail $\simeq t^{
-\mu-1}$ with $0<\mu<1$ and diverging mean waiting time, such as the ML PDF 
considered here, all tracers immobilize eventually with a long-time power-law
decay $\simeq t^{\mu-1}$ of the mobile mass. Our model with a ML PDF contains the
fractal MIM from \cite{schumer2003fractal} and the bi-fractional diffusion model 
from \cite{sandev2015distributed,maryshev2009non,chechkin2002retarding} for
specific choices of the scaling exponents as special cases. We find analytical
results up to the second moment in this special case that hold for all times. Our
ML model shows good fit results to the mobile mass decay of the MADE-1 experiment
\cite{adams1992field}. In addition we considered two extensions of the
immobilization time PDF $\gamma(\tau)$. First we introduced an exponential
tempering to analyze truncation effects. Second we considered a weighted sum of an
ML PDF and an exponential PDF. Both modifications yield even better fit results
than the ML model alone. While these extended forms involve additional model
parameters their better fit indicates that cut-offs in the power-law trapping
time density reflect better the physical situation, i.e., the system appears to
show finite size effects, similar to those obtained in trapping time PDFs in the
conductivity study \cite{edery2014origins}.

The ML model yields a good fit to the BTC ($R^2=0.990$) of tracers in karst aquifers
from \cite{goeppert2020experimental}. This allowed us to obtain model parameters
including the advection velocity and dispersion coefficient in the mobile zone.
Subsequently we calculated the moments of the mobile distribution and accounted for
time-dependent normalizations. We found temporally non-linear mass transport
$\langle x\rangle\simeq t^\mu$ and anomalous diffusion $\langle (x-\langle x
\rangle)^2\rangle\simeq t^{2\mu}$ in the long-time limit. Concurrently, the
concentration crossed over to a non-Gaussian immobile concentration. Mobile
tracers lead the immobile tracers in this long-time limit. We characterized the
non-Gaussianity using the skewness and kurtosis, of which the long-time limit
only depends on $\mu$, as shown in Appendix \ref{chkurtosis}.

Notably, a fit to the BTC with an exponential model matches the data quite reasonably
($R^2=0.940$) but yields an unlikely high diffusion constant. Nevertheless the mobile
concentration profiles appear reasonably similar for $t=$30 h and almost match for
$t=$200 h. The exponential model shows transient anomalous diffusion in this
time window, and the second central moment almost coincides with the ML model. It
is remarkable that the exponential and ML model have a fundamentally different
long-time behavior but yield very similar first, second, and second central
moments in the intermediate time-window, where the BTC measurements where taken.
In fact, our analysis demonstrates that from experimental data it is rather
tricky to distinguish even fundamentally different models based on transport
moments. The mobile mass, BTC, and concentration profile are much better suited
for this purpose. However, once fitted to the data, the moments demonstrate the
massively different transport efficiency at long times. Such knowledge is of
high relevance, e.g., to study the environmental impact of chemicals released
into rivers or aquifers. The existence of long retention times may be
underestimated by fits to exponential models and thus neglect potentially
dangerous leakage of chemicals at much longer times.

The situation is quite different in other systems in which more extensive data are
available, such as from simulations or single particle trapping experiments in live
biological cells or complex liquids. In such systems the moments can be efficiently
extracted and compared to different models. There, particles can undergo diffusion
with intermittent immobilization, as well. An example could be proteins diffusing
in the bulk cytoplasm of a live cell with intermittent binding to membrane receptors.
In fact three-dimensional trajectories of mRNA particles in yeast cells have been
observed to switch between diffusive, directed and confined motion as well as
becoming stationary \cite{thompson2010three}, similar to amoeboid motion on surfaces
\cite{beta}. Single molecule tracking of signaling proteins reveals intermittent
dynamics during which proteins effectively immobilize on activation
\cite{murakoshi2004single}. Membrane proteins and proteins in the cell nucleus have
been observed to split into mobile and immobile populations
\cite{manley2008high,kues2001visualization,weigel2013quantifying}.
We mention molecular dynamics
simulations of drug molecules in a water layer confined in a silica slit unveiling
intermittent immobilization due to surface adsorption with power-law distributed
trapping times \cite{fernandez2020diffusion}. Similar waiting time distributions
are observed in the short time motion of lipid granules in live yeast cells
\cite{lene}. In fact, for systems with power-law distributed immobilization times
or diffusion with strongly position-dependent diffusivity populations splitting
is a salient feature \cite{schulz2013aging,johannes1,andreypccp}.

The MIM can also be thought of as a special case of switching diffusion, when
a particle intermittently undergoes different modes of transport within a single
trajectory \cite{grebenkov2019TAMSD,tyagi2017non,baldovin2019polymerization}. 
When adding an advection-diffusion operator to the immobile concentration of EMIM
a switching diffusion process could be obtained.
In \cite{grebenkov2019TAMSD,tyagi2017non} a
single particle switches between states with different diffusivities with
fixed rates. If the observation time is small compared to
the mean residence time, transport anomalies arise. Examples for switching
diffusion include quantum dot tracers in the cytoplasm of mammalian cells which
switch between different mobilities \cite{sabri2020elucidating}. Molecular dynamics
simulations show that conformal changes of proteins induce fluctuations of the protein
diffusivity \cite{yumamoto2021universal}. A simple model of particles that can
aggregate and separate shows similar behavior \cite{hidalgo2020hitchhiker}. Polymers
change diffusivity during (de-)polymerization due to varying chain lengths leading
to transient non-Gaussian displacement PDFs \cite{baldovin2019polymerization}.
Switching behavior is also seen in potassium channels and nonintegrin receptors in
living cell membranes \cite{diego,carloprx} as well as for lipid motion in molecular
dynamics simulations of protein-crowded bilayer membranes \cite{ilpoprx}.
Similar population splitting is observed in the passive motion of tracers in mucin
gels \cite{mucussm,andreysm} or acetylcholine receptors in live cell membrane \cite{he}.
Population splitting into fractions with different diffusivities was also observed
for individually labeled lipids in the phospholipid membrane and of H-Ras proteins
at the plasma membrane \cite{schuetz1997single,lommerse2008single}.
Moreover, G-proteins have been observed to switch between states with different
diffusivities due to conformational changes and increased immobilization after
interaction \cite{sungkaworn2017single}. These cases of molecular transport can
represent scenarios in which our EMIM model or its extensions can provide
relevant insight into population splitting between mobile and immobile particle
fractions and their respective transport dynamics. Moreover, the breakthrough
curves discussed here can be used to deduce the first-passage dynamics to some
reaction center.

Our model is a starting point to describe molecular reactions of anomalously
diffusing tracers, such as reactions occurring in mobile and immobile zones in
rivers or reactions of molecules or tracers in biological cells. Recently,
reaction-subdiffusion systems have been analyzed using the Fokker-Planck-Kolmogorov
equation \cite{alexander2021reaction}. With our model it is possible to model
reactions that only occur when the particles immobilize and to find explicit
equations for the reaction products. We believe that the EMIM presented here
provides a flexible and unified description for mobile-immobile transport.

\begin{acknowledgments}
We acknowledge funding from the German Science Foundation (DFG gant ME 1535/12-1).
RM acknowledges the Foundation for Polish Science (Fundacja na rzecz Nauki Polskiej,
FNP) for funding within an Alexander von Humboldt Polish Honorary Research
Scholarship. AC acknowledgments the support of the Polish National Agency for
Academic Exchange (NAWA).
\end{acknowledgments}

\appendix

\section{Mobile mass using the ML trapping time PDF}
\label{chMML}

We calculate the mobile mass (\ref{eqMs}) for the concrete ML form of the trapping
time PDF. In Laplace space we find
\begin{eqnarray}
M_m(s)&=& \frac{M_0 }{s+\omega \beta \left[ 1-\frac{1}{1+\tau_\star^\mu s^\mu} \right]}\\
&=&\frac{M_0(1+\tau_\star^\mu s^\mu)}{s(1+\tau_\star^\mu s^\mu)+\omega\beta\tau_\star
^\mu s^\mu}\\
&=&\frac{M_0 (1+\tau_\star^\mu s^\mu+ \beta\omega\tau_\star^\mu s^{\mu-1})-M_0\beta
\omega\tau_\star^\mu s^{\mu-1}}{s(1+\tau_\star^\mu s^\mu +\beta\omega \tau_\star^\mu
s^{\mu-1})}\\
&=&\frac{M_0}{s}-M_0\beta\omega\tau_\star^\mu\frac{s^{\mu-2}}{\tau_\star^\mu s^\mu+\beta
\omega \tau_\star^\mu s^{\mu-1}+1}.
\label{eqMmcalc}
\end{eqnarray}
Now we use the geometric series for $s^{1-\mu}<\tau_\star^\mu s+\beta\omega\tau_\star
^\mu$
\begin{eqnarray}
\hspace{-1cm}\frac{1}{\tau_\star^\mu s^\mu+\beta\omega \tau_\star^\mu s^{\mu-1}+1}
&=&\frac{s^{1-\mu}}{\tau_\star^\mu s+\beta\omega\tau_\star^\mu}\cdot\frac{1}{1+\frac{
s^{1-\mu}}{\tau_\star^\mu s+\beta\omega \tau_\star^\mu}}\\
&=&\frac{s^{1-\mu}}{\tau_\star^\mu s+\beta\omega \tau_\star^\mu}\sum_{k=0}^\infty
(-1)^k\frac{s^{k(1-\mu)}}{(\tau_\star^\mu s+\beta\omega \tau_\star^\mu)^k}\\
&=&\sum_{k=0}^\infty\frac{(-1)^k}{\tau_\star^{\mu(1+k)}}\frac{s^{k(1-\mu)+(1-\mu)}}{
(s+\beta\omega)^{k+1}}\\
&=&\frac{1}{s\tau_\star^\mu(s+\beta\omega)}+\sum_{k=1}^\infty (-1)^k\frac{(-1)^k}{
\tau^{\mu(1+k)}}\frac{s^{k(1-\mu)+(1-\mu)}}{(s+\beta\omega)^{k+1}}.
\label{eqMmgeometric}
\end{eqnarray}
We use the Laplace inversion \cite[Eq. (2.5)]{prabhakar1971singular} and
(\ref{eqMmgeometric}) to transform (\ref{eqMmcalc}) to time-domain
\begin{eqnarray}
M_m(t)&=&M_0-M_0\beta\omega\tau_\star^\mu\left[\frac{1}{\beta\omega\tau_\star^\mu}(1-e^{
-\beta\omega t})\right.\nonumber \\
&&\hspace{4cm}\left.+M_0e^{-\beta\omega t}\sum_{k=1}^\infty(-1)^{k+1}\frac{t^{\mu
k+1}}{\tau^{\mu(k+1)}} E_{1,\mu k +2}^{k+1}(-\beta\omega t)\right]\nonumber\\
&=&M_0e^{-\beta\omega t}+N_0\beta\omega t\sum_{k=1}^\infty(-1)^{k+1}\left(\frac{t
}{\tau} \right)^{\mu k}E_{1,\mu k +2}^{k+1}(-\beta\omega t).
\label{eqMmappendix}
\end{eqnarray}
According to Gorenflo et al. \cite[Eq. (5.1.54)]{gorenflo2014mittag} $E_{1,\beta}^
\delta(z)=\frac{1}{\Gamma(\beta)}_1F_1(\delta,\beta,z)$ and thus (\ref{eqMmappendix}) 
\begin{equation}
M_m(t)=M_0 e^{-\beta \omega t} + M_0 \beta \omega t \sum_{k=1}^\infty 
\frac{(-1)^{k+1}}{\Gamma(\mu k+2)}\left(\frac{t}{\tau_\star}\right)^{\mu k}M(k+1,\mu k+2,
-\beta\omega t),
\end{equation}
where $_1F_1(a,b,z)\equiv M(a,b,z)$ is the Kummer function \cite{gorenflo2014mittag}.

\section{Long-time concentration profile}
\label{chConcentrationappendix}
\begin{figure}
\centering
\includegraphics{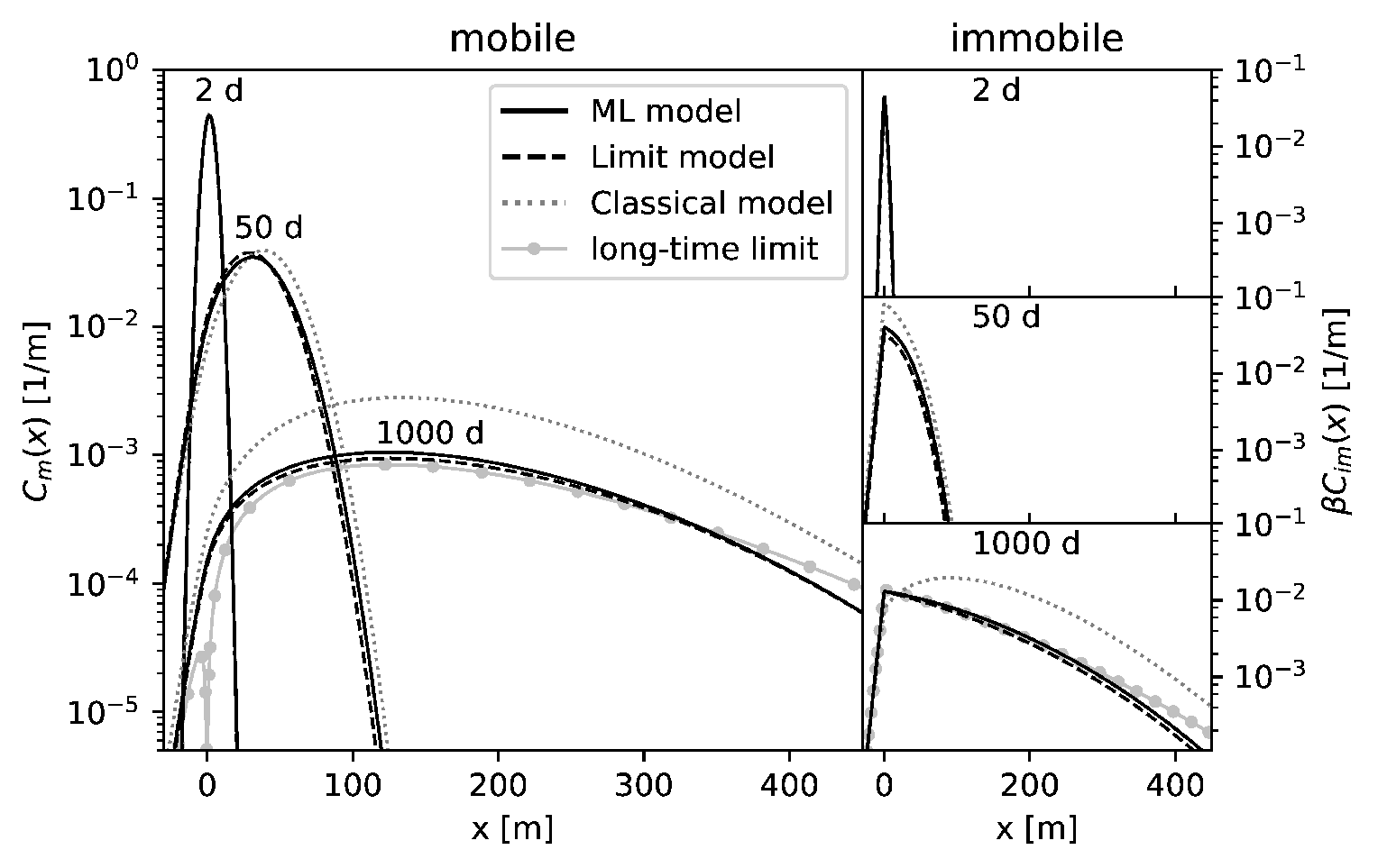}
\caption{Concentrations of mobile and immobile particles 2, 50 and 500 days after 
injection in the main and right panels, respectively. The parameters of all models
correspond to the fit to the mobile mass decay shown in Fig.~\ref{figmasscompare}.
At short times all models yield a Gaussian mobile distribution, which is increasingly
skewed for later times. Immobile particles have a non-Gaussian distribution at all
times. Cusps are clearly visible at $x=0$ and are a typical transport feature
for systems with diverging trapping times \cite{harvey,report}. The long-time
behaviors (\ref{eqCmlimappendix}) and (\ref{eqCimlimappendix}) are shown by the
grey lines with markers. The root at $x=0$ for the mobile concentration is an
artefact of our approximation, see text. We use 0.8 m/day and 4 $\mathrm{m}^2/$day
for $v$ and $D$, respectively. The parameters of the ML model can be found in
Tab.~\ref{tabPars}. The parameters for the classical and fractional model
correspond to a fit in \cite{schumer2003fractal} to the mobile mass decay of the
MADE-1 experiment, as can be seen in Fig.~\ref{figmasscompare} together with the
parameters \cite{schumer2003fractal}.}
\label{figConcentrationsApp}
\end{figure}

\textcolor{black}{Consider the fractal model.  We approximate $C_m(x,s)$
(\ref{eqCmxs}) using $\sqrt{1+z}\sim 1+z/2$ in the exponential and
$\sqrt{1+z}\sim 1$ in the first fraction for $z=4\phi(s)D/v^2\ll 1$ }
\begin{eqnarray}
C_m(x,s) &\overset{s\to 0}{\sim}&\frac{e^{\frac{vx}{2D}}}{v}e^{-\frac{v}{2D}(1+
\frac{2\phi(s)D}{v^2})|x|}\\
&\overset{s\to 0}{\sim}&\frac{1}{v}e^{\frac{v}{2D}(x-|x|)}e^{-\beta\omega\tau_\star^
\mu s^\mu\frac{|x|}{v}}
\label{eqAcmxs}
\end{eqnarray}
by using $\phi(s)=s+\beta\omega\tau_\star^\mu
s^\mu\approx\beta\omega\tau_\star^\mu\textcolor{black}{s^\mu}$ for small $s$. In
\cite[section 7.5]{gorenflo2014mittag} we find the Laplace transform pairs
\begin{eqnarray}
	\mathscr{L}^{-1}\{e^{-c s^\mu}\}&=& \frac{c\mu}{t^{\mu+1}}M_\mu(c t^{-\mu}) \label{eq01}\\
	\mathscr{L}^{-1}\{s^{\mu-1}e^{-c s^\mu}\}&=& \frac{1}{t^\mu}M_\mu(c t^{-\mu})\label{eq02}
\end{eqnarray}
for $c>0$ with the auxiliary function of Wright type,
\begin{equation}
M_\mu(z)=\sum_{n=0}^\infty \frac{(-z)^n}{n!\Gamma(-\mu n + (1-\mu))}.
\label{eqMwright}
\end{equation}
This yields the long-time limit of the mobile concentration,
\begin{equation}
C_m(x,t)\approx\frac{\beta\omega\tau_\star^\mu\mu}{v^2}e^{\frac{v}{2D}(x-|x|)}|x|t^{
-1-\mu}M_\mu\left(\frac{\beta \omega \tau^\mu_\star}{v} |x|t^{-\mu}\right),
\label{eqCmlimappendix}
\end{equation}
and the long-time limit of the immobile concentration using $C_{im}(x,s)=\omega
\frac{1-\gamma(s)}{s}C_m(x,s)$ with $\gamma(s)\approx1-\tau_\star^\mu s^\mu$,
\begin{equation}
C_{im}(x,t)\approx\frac{\beta\omega\tau_\star^\mu }{v}e^{\frac{v}{2D}(x-|x|)}t^{-\mu}
M_\mu\left(\frac{\beta\omega\tau^\mu_\star}{v}|x|t^{-\mu}\right).
\label{eqCimlimappendix}
\end{equation}
In Fig.~\ref{figConcentrationsApp} we show these approximations. Notice that at
$t=$1000 days, these approximations shown as a grey line with markers indeed
estimate quite well the results obtained from the inverse Laplace inversion.
Note also how the factor $|x|$ and finite value of $M_\mu(0)=1/\Gamma(1-\mu)$ leads
to a dip to zero at $x=0$ for this approximation. This is an artefact of our
approximation of $C_m(x,s)$ (\ref{eqAcmxs}), as it does not depend on $s$ for $x=0$.

\section{Additional plots of moments}

\begin{figure}
\includegraphics{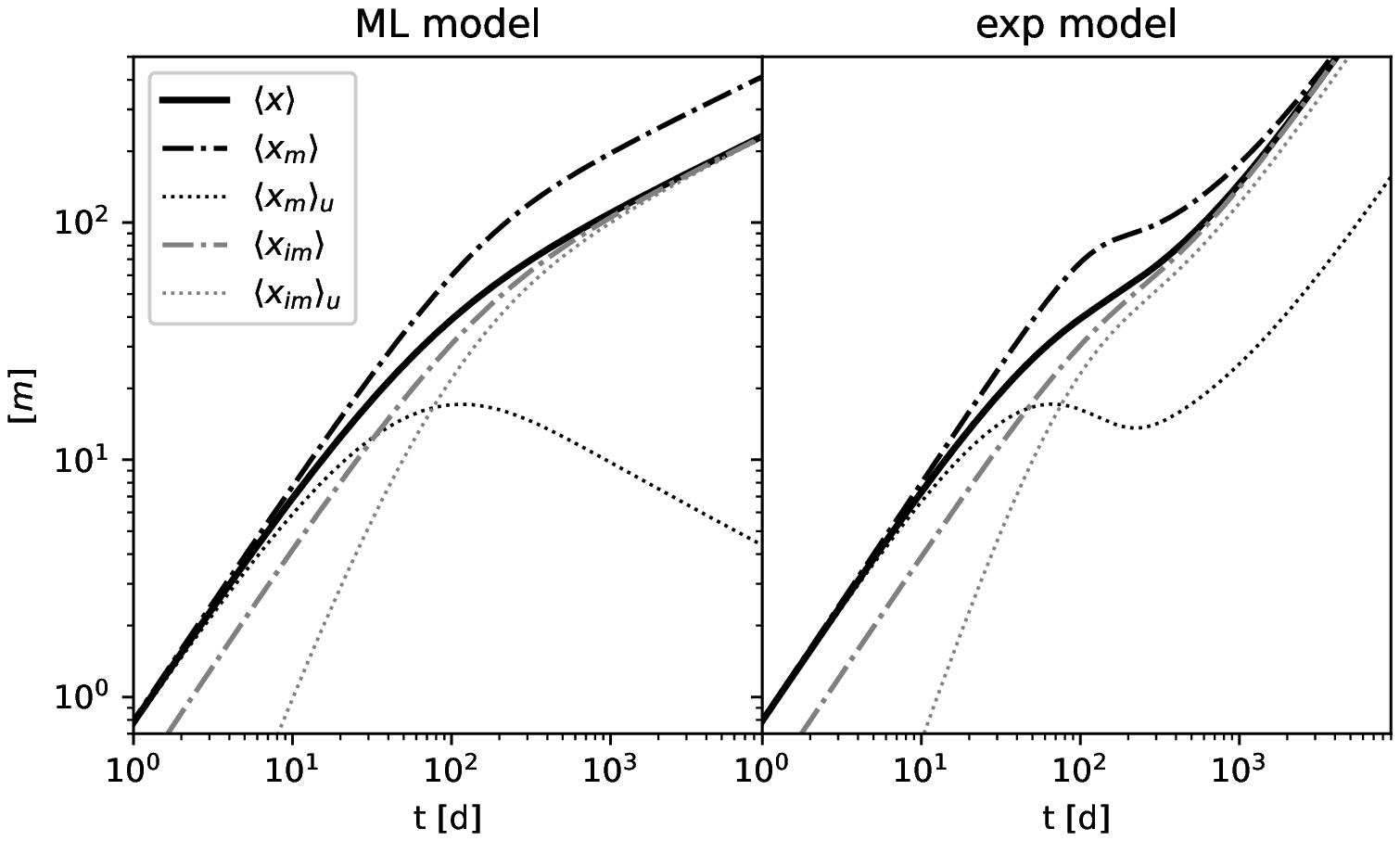}
\includegraphics{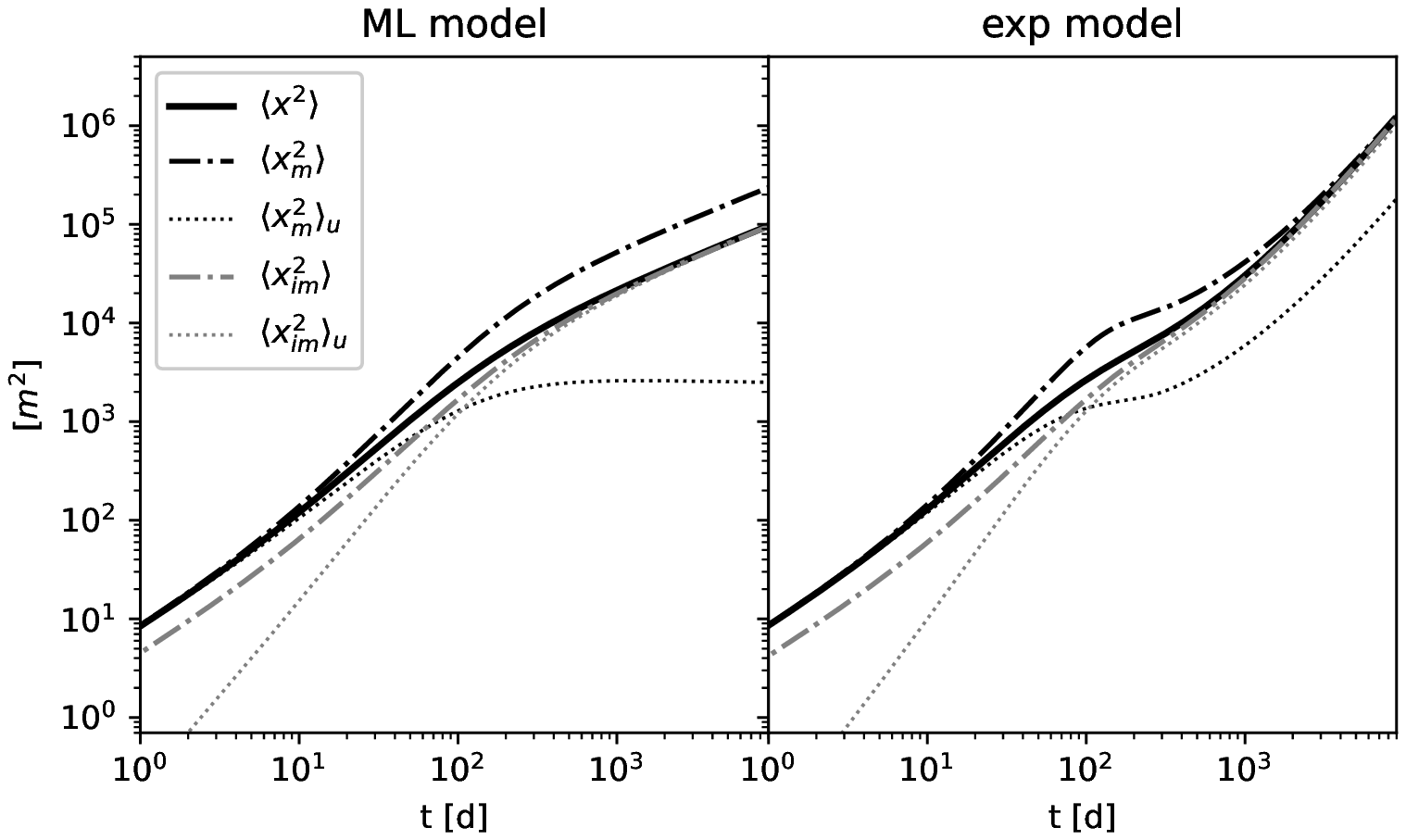}
\caption{Comparison of the normalized ($\langle x^n_{m,im}\rangle$), unnormalized
($\langle x_{m,im}\rangle_u$) and total $n$-th order moment $\langle x^n\rangle$.
The upper panel shows $n=1$ and the lower $n=2$ with double-logarithmic scales.
The left panel shows the ML model and the right one shows the exponential model.
In both panels, both $\langle x_m\rangle_u$ and $\langle x_m\rangle$ are a good
approximation for $\langle x\rangle$ for short times, due to the high mobile
fraction.  The same holds for long times for the immobile moment. In the ML model,
the first moment of the mobile particles has the same power-law behavior $\propto
t^\mu$ as the immobile tracers but with a larger coefficient. The parameters are
the same as in Fig.~\ref{figmasscompare}a and are given in Tab.~\ref{tabPars}.}
\label{figxallAppendix}
\end{figure}

We show additional plots of the first and second moments in Fig.~\ref{figxallAppendix}.
The left panels show our model using an ML trapping time PDF and the right panels
using an exponential trapping time PDF. The former demonstrates a transition from
normal Brownian to anomalous behavior.

\section{Skewness and Kurtosis}
\label{chkurtosis}

\begin{figure}
\includegraphics{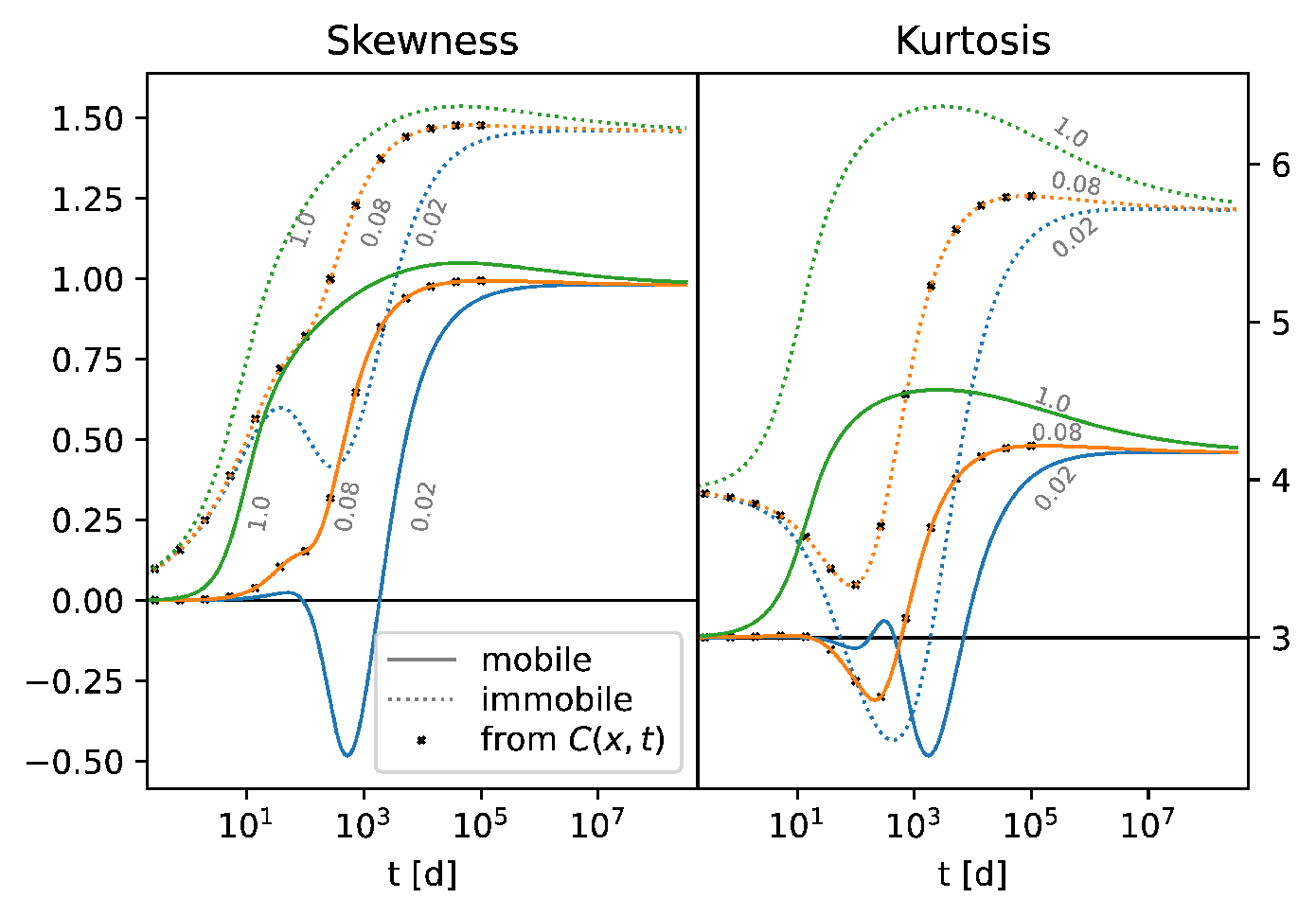}
\caption{Skewness and kurtosis for the ML model. Limits are calculated with the
derivatives (\ref{eqx3scaled}) in Fourier-space. Small numbers next to the lines
indicate the values
of $\beta$. Crosses mark the corresponding values for $\beta=0.08$ obtained from
the plume profile (\ref{eqCmxs}) after numerical Laplace inversion. The remaining
parameters are the same as in Fig.~\ref{figmasscompare}a, i.e., we have $\mu=0.33$,
$\omega=$1/day, and $\tau_\star$=1 day.}
\label{figkurtosis}
\end{figure}

Similarly to (\ref{eqx2scaled}) we calculate the third and fourth unnormalized moments
\begin{eqnarray}
\hspace{-1cm}\langle x_m^3(s)\rangle_u=\left.-i\frac{\partial^3}{\partial k^3}
\frac{C_m(k,s)}{M_0}\right|_{k=0}&=\frac{12 D v}{\left(s+\beta\omega[1-\gamma(s)]
\right)^3}+\frac{6 v^3}{\left(s+\beta\omega[1-\gamma(s)]\right)^4}\nonumber\\
\hspace{-1cm}\langle x_m^4(s)\rangle_u=\left.\frac{\partial^4}{\partial k^4}
\frac{C_m(k,s)}{M_0}\right|_{k=0}&=\frac{24 D^2}{(s+\beta\omega[1-\gamma(s)])^3}
+\frac{72 D v^2}{(s+\beta \omega[1- \gamma(s)])^4}+\frac{24 v^4}{(s+\beta\omega[1-
\gamma(s)])^5}.
\label{eqx3scaled}
\end{eqnarray}
When dividing by the fraction of mobile mass, we can calculate the skewness
\begin{equation}
\tilde\mu_{3,m}=\left\langle\left(\frac{x_m-\langle x_m\rangle}{\sqrt{\langle(
x_m-\langle x_m\rangle)^2\rangle}} \right)^3\right\rangle
\label{eqskew}
\end{equation}
and kurtosis
\begin{equation}
\tilde\mu_{4,m}=\left\langle\left(\frac{x_m-\langle x_m\rangle}{\sqrt{\langle(
x_m-\langle x_m\rangle)^2\rangle}} \right)^4\right\rangle
\label{eqkurt}
\end{equation}
of the mobile solutes. Using (\ref{eqCimks}), we obtain the skewness and kurtosis
for the immobile plume.

A normal distribution in one dimension has skewness zero and kurtosis three.
Deviations from these values characterize non-Gaussianity. Fig.~\ref{figkurtosis}
shows the skewness and kurtosis using the ML model for different values of $\beta$
for $\omega\tau_\star=1$. The mobile plume shows no initial skewness, as expected by
the short-term Gaussian distribution. Small values of $\beta$ yield negative
skewness for intermediate time scales, i.e., a leading edge of the mobile plume
profile. In \cite{schumer2003fractal} negative skewness is found for small $\beta$,
as well. We find that the long-time limit of the skewness is independent of $\beta$
and positive for $\mu\lessapprox 0.73$. In addition, we numerically find positive
skewness for intermediate times when decreasing $v$ and leaving all remaining
parameters constant for $\mu=0.33$ and $\beta=0.01$.

The long-time limit is independent of $\beta$ and positive for the chosen $\mu=0.33$.
This corresponds to a leading tail of the mobile plume profile. The immobile
distribution has positive skewness at all time scales and is non-monotonous for
$\beta=0.02$. To verify our results, we additionally calculate the skewness and
kurtosis from the plume profile (\ref{eqCmxs}) for $\beta=0.08$. The resulting
crosses in Fig.~\ref{figkurtosis} show good agreement.

The kurtosis measures how much of a distribution is concentrated in the tails. 
As Fig.~\ref{figkurtosis} shows, the mobile distribution starts at three and 
has minima at intermediate times below this value for $\beta=0.02$ and $\beta=0.08$.

\begin{figure}
\includegraphics{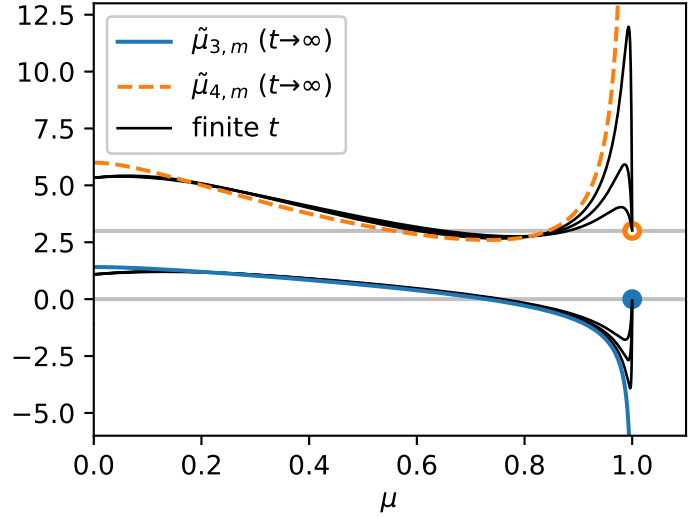}
\caption{Long-time limits of the skewness (\ref{eqskew}) and kurtosis (\ref{eqkurt})
that only depend on $\mu$ and the sign of $v$ (chosen positive here) for the ML and
fractal models. The expressions for $\mu<1$ diverge for $\mu\to 1$. For $\mu=1$, we
have $\tilde \mu_{3,m}=0$ and $\tilde \mu_{4,m}=1$. Grey lines indicate these values.
Black solid lines indicate the kurtosis for finite $t=500,200,100$ from top to bottom. 
They all reach the Gaussian value of three at $\mu=1$ and are continuous. The skewness
for finite $t=500,1000,2000$ show black solid lines ending at zero for $\mu=1$. They
are continuous, as well. Black lines were obtained from the asymptotic expressions
of the moments for $\beta=v=D=1$.}
\label{figskewlimit}
\end{figure}

We calculate the long-time limits
\begin{equation}
\langle x_m^3\rangle\approx\frac{\frac{12Dvt^{3\mu-1}}{\beta^3\Gamma(3\mu)}+\frac{
6v^3t^{4\mu-1}}{\beta^4\Gamma(4\mu)}}{\frac{t^{\mu-1}}{\beta\Gamma(\mu)}}	
\label{eqx3lim}
\end{equation}
and 
\begin{equation}
\langle x^4_m\rangle\approx\frac{\frac{24D^2t^{3\mu-1}}{\beta^3\Gamma(3\mu)}+\frac{
72Dv^2t^{4\mu-1}}{\beta^4\Gamma(4\mu)}+\frac{24v^4t^{5\mu-1}}{\beta^5\Gamma(5\mu)}}{
\frac{t^{\mu-1}}{\beta\Gamma(\mu)}}
\label{eqx4lim}
\end{equation}
by using the Tauberian theorem. These results match earlier results found in
\cite{zhang2008moments}. We plug these into (\ref{eqskew}) and (\ref{eqkurt}).
For $\mu=1$ we find $\tilde \mu_3=0$ and $\tilde \mu_4=3$, which matches the
normal distribution for $\mu=1$ found by \cite{schumer2003fractal}.
The long-time limits of the skewness only depends on $\mu$ and the sign of $v$,
which we assume to be positive, here
\begin{eqnarray}
\hspace{-2cm}\lim_{t\to\infty}\tilde\mu_3&=-\frac{2\sqrt{\frac{\Gamma(3\mu)}{\Gamma
(\mu)}}\left(12\sqrt{\pi}\Gamma(\mu)\Gamma(2\mu)^3\Gamma(4\mu)-12\sqrt{\pi}\Gamma
(2\mu)^4\Gamma(3\mu)\right)}{\left(2\Gamma(2\mu)^2-\Gamma(\mu)\Gamma(3\mu)\right)
^{3/2}\left(4\sqrt{\pi}\Gamma(2\mu)\Gamma(4\mu)-16^{\mu}\Gamma(\mu)\Gamma(3\mu)
\Gamma\left(2\mu+\frac{1}{2}\right)\right)}\nonumber\\
&-\frac{2\sqrt{\frac{\Gamma(3\mu)}{\Gamma(\mu)}}\left(16^{\mu}\Gamma(\mu)^3\Gamma
(3\mu)^2\Gamma\left(2\mu+\frac{1}{2}\right)-4\sqrt{\pi}\Gamma(\mu)^2\Gamma(2\mu)
\Gamma(3\mu)\Gamma(4\mu)\right)}{\left(2\Gamma(2\mu)^2-\Gamma(\mu)\Gamma(3\mu)
\right)^{3/2}\left(4\sqrt{\pi}\Gamma(2\mu)\Gamma(4\mu)-16^{\mu}\Gamma(\mu)\Gamma
(3\mu)\Gamma\left(2\mu+\frac{1}{2}\right)\right)}\nonumber\\
&-\frac{2\sqrt{\frac{\Gamma(3\mu)}{\Gamma(\mu)}}\left(-3\Gamma(\mu)\Gamma(2\mu)^2
\Gamma(3\mu)\left(16^{\mu}\Gamma(\mu)\Gamma\left(2\mu+\frac{1}{2}\right)-2\sqrt{\pi}
\Gamma(3\mu)\right)\right)}{\left(2\Gamma(2\mu)^2-\Gamma(\mu)\Gamma(3\mu)\right)^{
3/2}\left(4\sqrt{\pi}\Gamma(2\mu)\Gamma(4\mu)-16^{\mu}\Gamma(\mu)\Gamma(3\mu)\Gamma
\left(2\mu+\frac{1}{2}\right)\right)}
\label{eqskewlim}
\end{eqnarray}
while the long-time limit of the kurtosis only depends on $\mu$
\begin{eqnarray}
\hspace{-2cm}\lim_{t\to\infty}\tilde\mu_4&=-\frac{3\Gamma(3\mu)\left(8\Gamma(\mu)
\Gamma(2\mu)^3\Gamma(3\mu)\Gamma(5\mu)-8\Gamma(2\mu)^4\Gamma(3\mu)\Gamma(4\mu)
\right)}{\Gamma(\mu)\left(\Gamma(\mu)\Gamma(3\mu)-2\Gamma(2\mu)^2\right)^2\Gamma
(4\mu)\Gamma(5\mu)}\nonumber\\
&-\frac{3\Gamma(3\mu)\left(\Gamma(3\mu)\Gamma(4\mu)\Gamma(5\mu)\Gamma(\mu)^3+4
\Gamma(2\mu)^2\Gamma(4\mu)\Gamma(5\mu)\Gamma(\mu)^2\right)}{\Gamma(\mu)\left(\Gamma
(\mu)\Gamma(3\mu)-2\Gamma(2\mu)^2\right)^2\Gamma(4\mu)\Gamma(5\mu)}.
\label{eqkurtlim}
\end{eqnarray}
Fig.~\ref{figskewlimit} shows the limiting values. The skewness takes positive values
for $\mu<0.73$, i.e. the mobile plume has a leading tail, and negative values
otherwise. The kurtosis is always higher than three except for $0.56<\mu<0.84$,
meaning that for $0.56<\mu<0.84$ more mobile particles are within the standard
deviation than for a normal distribution and thus effect a pronouncedly
non-Gaussian distribution.

We note the apparent discontinuity of the long-time limits of the skewness
(\ref{eqskewlim}) and kurtosis (\ref{eqkurtlim}), as shown in Fig.~\ref{figskewlimit}.
Note that we get finite values of the kurtosis for finite $t$, as evidenced by
Fig.~\ref{figskewlimit}. This property will be analyzed in more detail elsewhere.

\section{Simulation}
\label{chSim}

\begin{figure}
\includegraphics{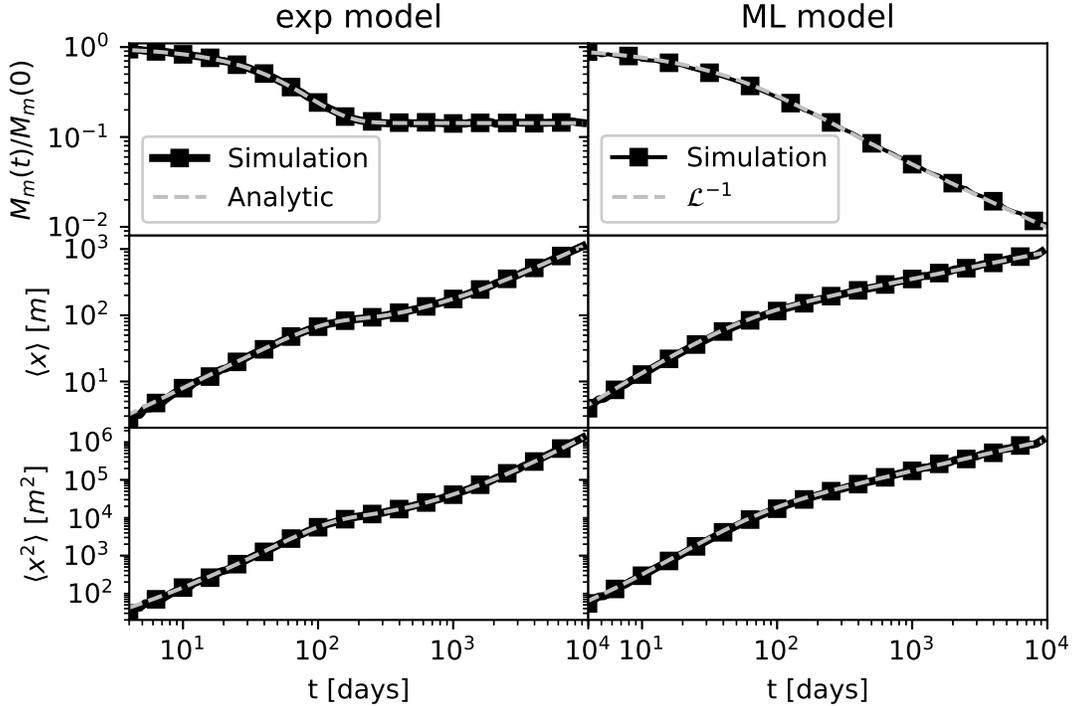}
\caption{Validation of our results with particle-tracking simulations. Square
symbols denote values obtained from simulations using the method proposed in
\cite{benson2019random}. For the exponential model we use our analytical results
and for the  ML model we compare it to Laplace inversions of (\ref{eqxscaled})
and (\ref{eqx2scaled}). Parameters are the same as in Fig.~\ref{figmasscompare}a.}
\label{figsim}
\end{figure}

We implement a particle-tracking simulation using the "space-domain method"
\cite{benson2019random}, in which
the particle makes a jump $\Delta x$ drawn from the jump length PDF $\lambda(x)$ in
the fixed time $\Delta t$ \cite{benson2019random}. After each jump the particle
immobilizes for a duration drawn from $\psi(t)$ with probability $1-\exp(-\omega
\beta\Delta t)$. For a waiting time PDF with the tail $\psi(t)\propto 1/t^{1+\mu}$
we use the method proposed by Kleinhans and Friedrich \cite{kleinhans2007CTRW}.
Results for the mobile mass, the first and second moment are shown in Fig.~\ref{figsim}.

\end{document}